\documentclass[pre,a4paper,superscriptaddress,showpacs,showkeys]{revtex4-1}
\pdfoutput=1
\usepackage{amsmath,amsfonts,amssymb,bm,color,graphicx}
\usepackage{epsfig,grffile,tipa}
\usepackage{hyperref}
\renewcommand{\vec}[1]{\bm{#1}}
\renewcommand{\tensor}[1]{\bm{#1}}

\newcommand{\uvec}[1]{\hat{\bm{#1}}}
\newcommand{\body}[1]{\widetilde{#1}}
\newcommand{\vbody}[1]{\widetilde{\bm{#1}}}
\newcommand{\df}[1]{\text{d}#1\,}
\newcommand{\dt}[1]{\dot{#1}}

\newcommand{\vdf}[1]{\text{d}\vec{#1}\,}
\newcommand{\avg}[1]{\left\langle#1\right\rangle}
\newcommand{\norm}[1]{\left\lvert#1\right\lvert}

\DeclareGraphicsExtensions{{.pdf}}
\begin{document}
 \title{Hydrodynamic Interactions of Self-Propelled Swimmers}
 \author{John J. Molina} \email{john@cheme.kyoto-u.ac.jp}
 \affiliation{Department of Chemical Engineering, Kyoto University,
   Kyoto 615-8510} \author{Yasuya Nakayama} \affiliation{Department of
   Chemical Engineering, Kyushu University, Fukuoka 819-0395}
 \author{Ryoichi Yamamoto} \email{ryoichi@cheme.kyoto-u.ac.jp}
 \affiliation{Department of Chemical Engineering, Kyoto University,
   Kyoto 615-8510} \date{\today}
 \begin{abstract}
   The hydrodynamic interactions of a suspension of self-propelled
   particles are studied using a direct numerical simulation method
   which simultaneously solves for the host fluid and the swimming
   particles. A modified version of the ``Smoothed Profile'' method
   (SPM) is developed to simulate microswimmers as squirmers, which
   are spherical particles with a specified surface-tangential slip
   velocity between the particles and the fluid. This simplified
   swimming model allows one to represent different types of
   propulsion (pullers and pushers) and is thus ideal to study the
   hydrodynamic interactions among swimmers. We use the SPM to study
   the diffusive behavior which arises due to the swimming motion of
   the particles, and show that there are two basic mechanisms
   responsible for this phenomena: the hydrodynamic interactions
   caused by the squirming motion of the particles, and the
   particle-particle collisions. This dual nature gives rise to two
   distinct time- and length- scales, and thus to two diffusion
   coefficients, which we obtain by a suitable analysis of the
   swimming motion. We show that the collisions between swimmers can
   be interpreted in terms of binary collisions, in which the
   effective collision radius is reduced due to the collision dynamics
   of swimming particles in viscous fluids. At short time-scales, the
   dynamics of the swimmer is analogous to that of an inert tracer
   particle in a swimming suspension, in which the diffusive motion is
   caused by fluid-particle \textit{collisions}. Our results, along
   with the simulation method we have introduced, will allow us to
   gain a better understanding of the complex hydrodynamic
   interactions of self-propelled swimmers.
 \end{abstract}
\maketitle

\section{Introduction}
Swimming microorganisms, from bacteria, to algal cells, to
spermatozoa, are ubiquitous in biological processes, and even though
the specific propulsion mechanism can vary, the motion of these
microswimmers is defined by two basic characteristics: (1) they are
moving at very low Reynolds number, where viscous forces are dominant
(inertial forces can be neglected), and thus (2) the net force on the
body (including the cilia or flagella used to generate the motion) is
zero\cite{PURCELL:1977tk,Lauga:2009ku}. Although we have a clear
understanding of how these organisms can generate motion, the physical
properties of a suspension of such swimmers are still not completely
understood\cite{Manghi:2006jx,Golestanian:2011fw,Guasto:2012uj,Ramaswamy:2010bf,Koch:2011ko}.
In particular, the rheological properties show a non-trivial
dependence on the concentration of
swimmers\cite{Ishikawa:2007gk,Sokolov:2009dy,Rafai:2010cn}, which
deviates considerably from that of inert force-free colloidal
particles\cite{GK:1970wl}. In addition, hydrodynamic interactions
among swimmers have also been shown to give rise to collective
motion\cite{Sokolov:2007td,Drescher:2009uj}. However, theoretical
studies on such phenomena have either neglected the role of
hydrodynamic interactions, or used only far-field
approximations\cite{Toner:2005ty,Koch:2011ko}.

In addition to the biological interest of understanding the
hydrodynamic interactions of swimming microorganisms, recent
experimental work has shown how to design micro-motors which can swim
in the absence of external fields thanks to interfacial
\textit{phoretic}
effects\cite{Anderson:1989ua,Paxton:2006dm,Sen:2005wq,Howse:2007ed,Jiang:2010el},
which allow the particles to transform the local chemical or thermal
energy into mechanical energy. Furthermore, the possibility of using
these systems to perform work, whether it be to transport
cargo\cite{Baraban:2012jf,Kolmakov:2011ba} or separate particle
mixtures\cite{Yang:2012eb}, has led to many recent experimental and
simulation studies. Although a simple theoretical framework for these
swimming systems, which can be used to study the effect of the shape
and surface properties on the swimming motion, has already been
proposed\cite{Golestanian:2007hu,Brady:2011bh}, the collective
dynamics, and in particular the role of the hydrodynamic interactions
remains an open question. In this paper, we present a direct numerical
simulation method (DNS) for self-propelled particles which attempts to
overcome this difficulty, by solving the equations of motion for the
host fluid as well as the swimming particles.

A detailed description of the swimming mechanism of real
microorganisms remains computationally prohibitive, particularly if
one is interested in the collective dynamics of such systems;
therefore, we must look for a simple generic swimmer, which can be
easily modeled, but which manages to reproduce the basic flow
properties of actual swimmers, such that the hydrodynamic interactions
are accurately reproduced. The ``squirmer'' model we use was
introduced over 50 years ago\cite{MJ:1969wg,Blake:2006ig} to study the
propulsion of spherical ciliate particles, which generate motion
through the synchronized beating of an envelope of cilia at their
surface. Fortunately, on a mesoscopic scale, the effect of this
surface motion can be replaced by a specified slip velocity between
the particle and the fluid. In addition, the parameters of the model
can be tuned to mimic different propulsion mechanism, allowing us to
study the hydrodynamic interactions of different (idealized)
microswimmers within a unified framework. We present a modified
``Smoothed Profile'' method\cite{Nakayama2005a}, previously developed to
study colloidal dispersions, which is capable of incorporating this
squirming motion by imposing the appropriate slip boundary conditions
at the surface of the particles. The advantage of this method is that
it allows us to accurately and efficiently include the many-body
hydrodynamic interactions among the particles, and it can be
easily extended to complex host fluids.

In this work we study the effect of hydrodynamic interactions on the
dynamics of a suspension of self-propelled particles. The swimming
motion of the particles is known to give rise to a diffusive
behavior\cite{Ishikawa:2007cf}, even in the absence of thermal
fluctuations, and a tracer particle placed in such a system will also
undergo diffusion\cite{Kim:2004bq,Leptos:2009kd}. However, the nature
of the diffusion is different in both cases, as is evidenced by the
different scaling behavior with the concentration of
swimmers\cite{Ishikawa:2010ho}. In the former case, the diffusion is
caused primarily by particle-particle collisions, while the latter is
due to the hydrodynamic interactions caused by the squirming motion of
the particles. By a suitable analysis of the particle displacements,
and the decay in the velocity fluctuations, we show that both
phenomena are present in the motion of the squirmers themselves, as
should be expected. We are thus able to extract the two underlying
time- and length-scales of the system, corresponding to the two
distinct diffusive motions, just from the particle trajectories. The
importance of this cannot be understated, as these two mechanisms
(hydrodynamics and particle collisions) are known to be fundamental in
determining the physical properties of swimming
suspensions\cite{Sokolov:2012td}; additionally, measuring tracer
diffusion in dense suspensions experimentally can be quite
challenging, and the available simulation methods for this are
somewhat involved\cite{Ishikawa:2010ho}. Finally, using basic results
from kinetic theory, we analyze the collisions of the swimmers to
derive effective collision radii, which are shown to be independent of
concentration, and which depend only on the squirming mode of the
swimmers (i.e., the strength of the pusher/puller character). Somewhat
surprising is the fact that these effective particle sizes are
considerably smaller than the actual size of the particles, which is a
consequence of the collision dynamics of swimmers in viscous fluids.
This paper is organized as follows: Section~\ref{s:mm} introduces the
mathematical model of the swimmers and the simulation method, Section~\ref{s:test} validates
the computational method against known analytical results, and
Section~\ref{s:main} presents the results of our study.
\section{The Model and Methods}
\label{s:mm}
\subsection{Swimmer Model: Blake's Squirmers}
\label{s:blake}
\begin{figure}[ht!]
  \centering
  \includegraphics[width=0.2\textwidth]{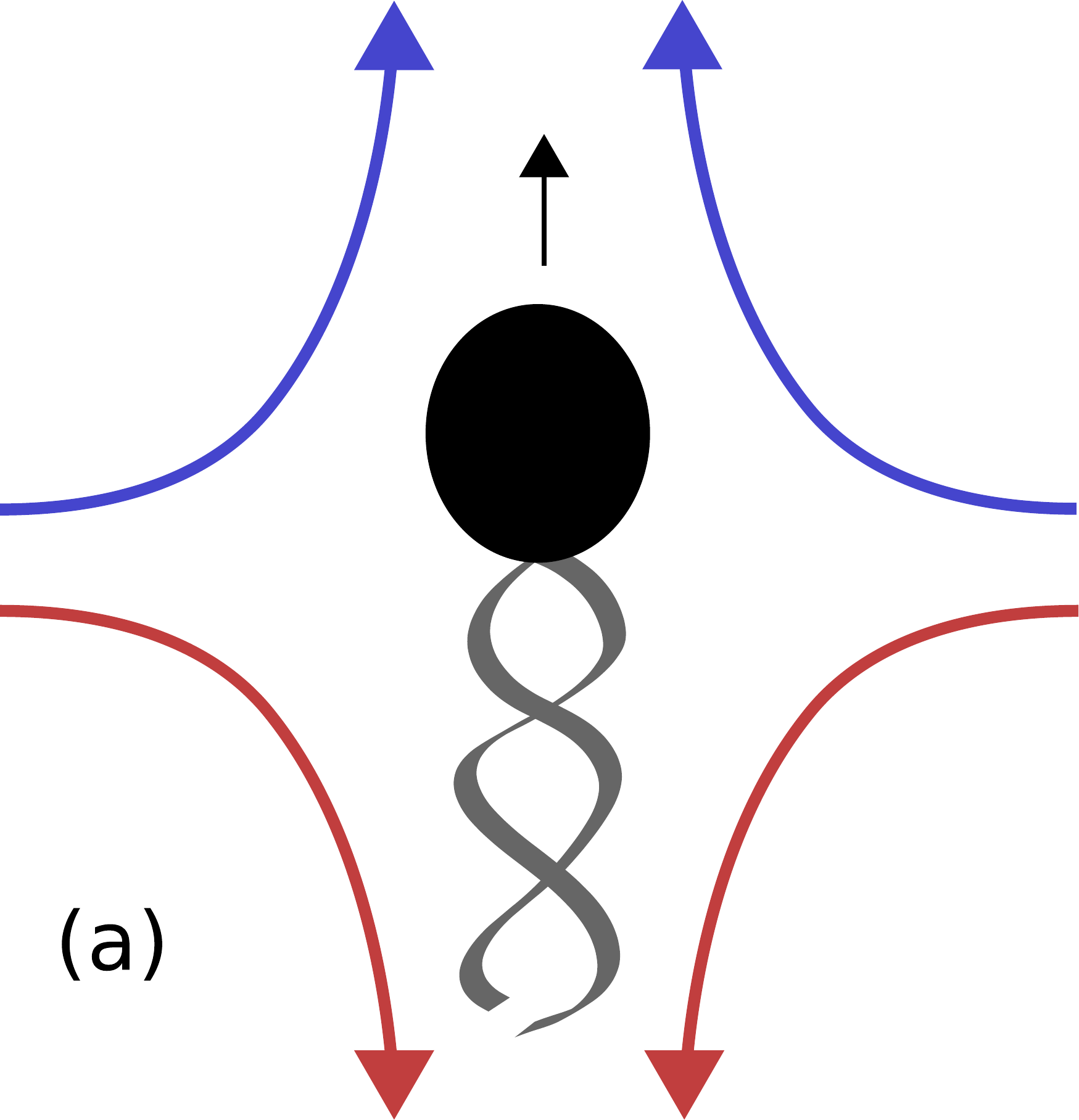}\quad%
  \includegraphics[width=0.2\textwidth]{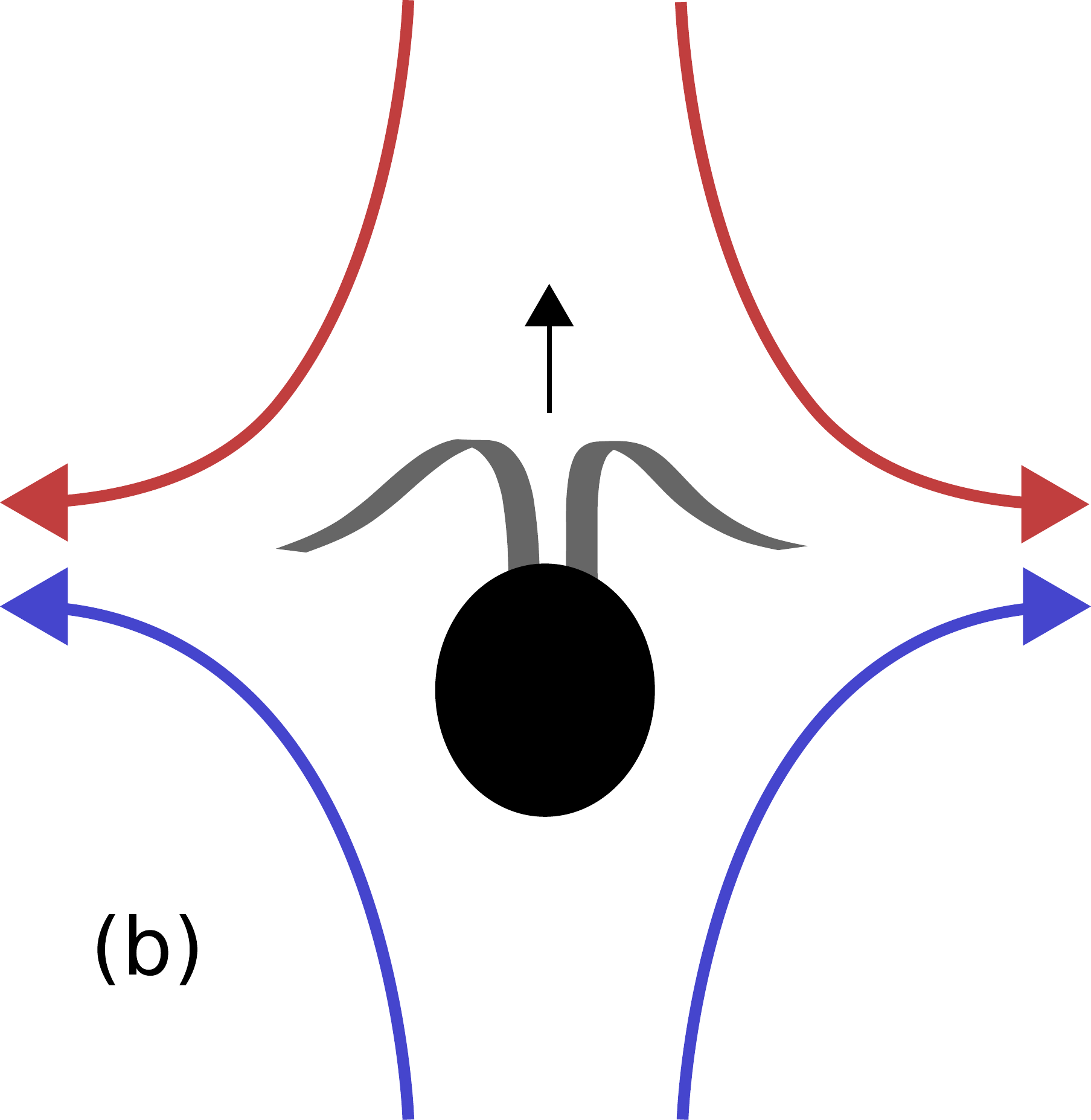}\\
  \vspace{0.25cm}
  \includegraphics[width=0.2\textwidth]{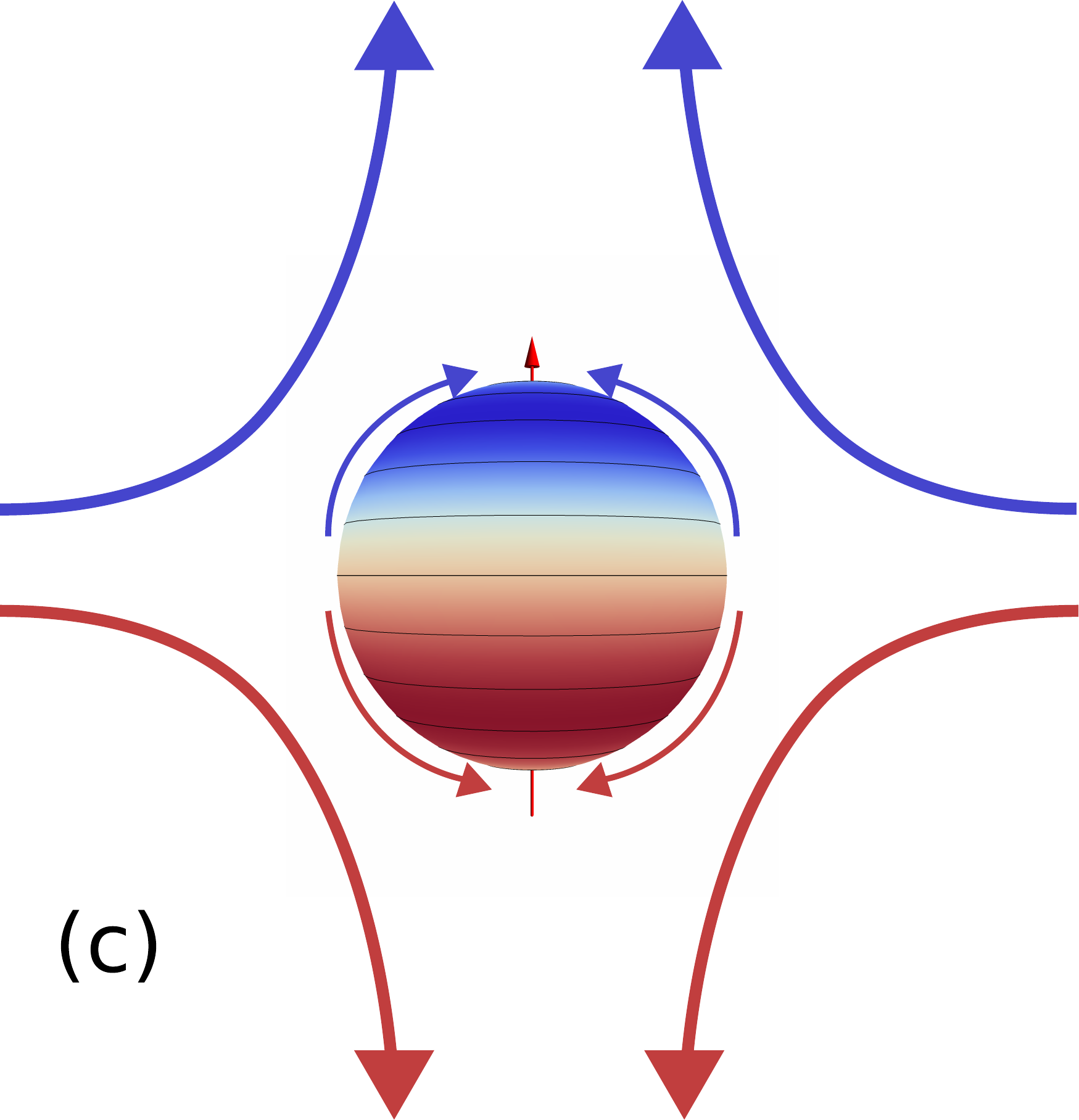}\quad%
  \includegraphics[width=0.2\textwidth]{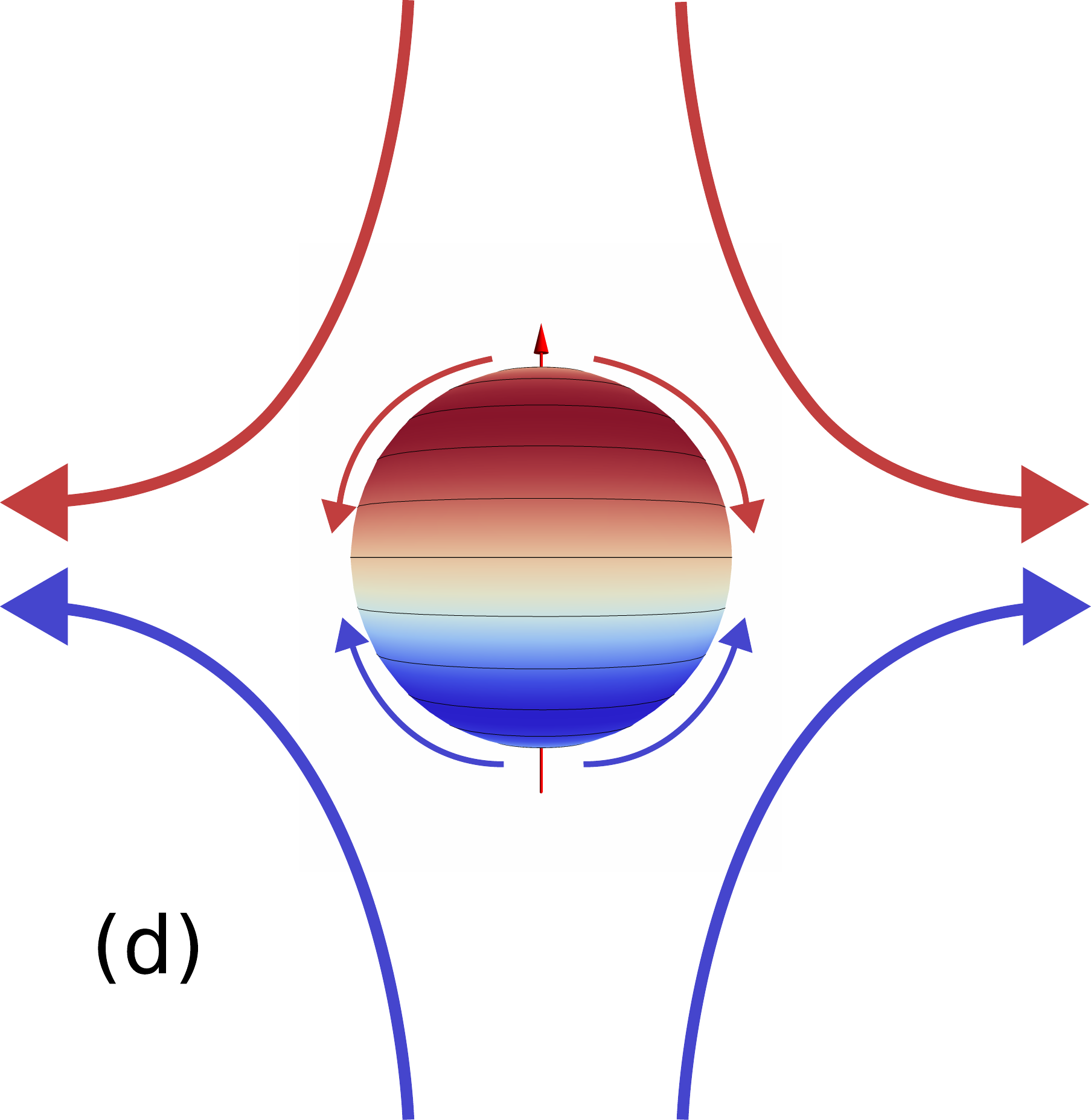}
  \caption{\label{f:pushpull}(color online) Schematic representation
    of the propulsion mechanism and flow profiles of a pusher and a
    puller, (a) and (b) respectively. These swimmers can be
    represented using Blake's squirming model, in which the detailed
    propulsion mechanism is replaced by a specified slip velocity at
    the surface of the particles, (c) and (d), for pushers and
    pullers, respectively.}
\end{figure}
We consider a simple model of self-propelled spherical swimmers,
originally introduced by Lighthill\cite{MJ:1969wg} and later extended
by Blake\cite{Blake:2006ig}, which move due to a self-generated
surface-tangential velocity $\vec{u}^s$. This specific mechanism was
proposed as a model for an ideal ciliate particle, in which the
synchronized beating of the cilia at the surface gives rise to a net
motion, in the absence of any external fields. If one assumes that the
displacements of this envelope of cilia are purely tangential, then
the effective (time-averaged) slip velocity for these
\textit{squirmers} is given by\cite{Blake:2006ig}
\begin{equation}\label{e:squirm_full}
  \vec{u}^s(\uvec{r}) = \sum_{n=1}^\infty\frac{2}{n\left(n+1\right)}
  B_n \left(\uvec{e}\cdot\uvec{r}\uvec{r} -
    \uvec{e}\right)P_n^\prime\left(\uvec{e}\cdot\uvec{r}\right)
\end{equation}
where $\uvec{e}$ is the squirmer's fixed swimming axis (i.e., we consider
that each squirmer carries with it a fixed coordinate system which
determines its preferred swimming direction at each instant),
$\uvec{r}$ is a unit vector from the particle center to a point on the
surface, $P_n^\prime$ is the derivative of the $n$-th order Legendre
polynomial, and $B_n$ is the amplitude of the corresponding mode.
Neglecting all squirming modes higher than three, $B_n = 0 \,(n\ge
3)$, the following simple expression for the surface
tangential velocity, as a function of the polar angle $\theta =
\cos^{-1}{\left(\uvec{r}\cdot\uvec{e}\right)}$, is obtained
\begin{equation}\label{e:squirm}
  \vec{u}^s(\theta) = B_1\left(\sin{\theta} +
    \frac{\alpha}{2}\sin{2\theta}\right)
\end{equation}
where $\alpha=B_2/B_1$ determines whether the swimmer is a pusher
($\alpha < 0$) or a puller ($\alpha > 0$). An example of the former
are spermatozoa and most bacteria, of the latter the unicellular alga
\textit{Chlamydomonas}. A schematic representation of the flow profile
generated by these two types of swimmers is given in
Figure~\ref{f:pushpull}. Although the squirmer model we adopt forgoes
describing the detailed propulsion mechanism, it is capable of
distinguishing between pushers/pullers and provides an adequate
approximation to the far-field flow profile generated by these
swimmers. For Newtonian fluids, which is the only case considered here,
the swimming speed $U$ of the squirmer is determined uniquely by the
first mode $B_1$, irrespective of the size of the particle, as $U=2/3
B_1$, while the second mode gives the strength of the
stresslet\cite{Ishikawa:2006hf,Zhu:2012ht}. The velocity field
generated by a single such squirmer, in the Stokes regime, was solved
analytically by Ishikawa et al.\cite{Ishikawa:2006hf} and is given, in
the laboratory frame (fluid at rest far away from the particle), by
 \begin{equation}\label{e:squirm_field}
   \vec{u}(\vec{r}) = B_1\frac{a^2}{r^2}\left[
     \frac{a}{r}\left(\frac{2}{3}\uvec{e} +
       \sin{\theta}\,\uvec{\theta}\right)
     + \frac{\alpha}{2}\left\{
       \left(\frac{a^2}{r^2} - 1\right)\left(3\cos^2{\theta} -
         1\right)\uvec{r}
       +\frac{a^2}{r^2}\sin{2\theta}\,\uvec{\theta}
     \right\}
   \right]
\end{equation}
where $a$ is the radius of the particle. Notice that for neutral
swimmers ($\alpha=0$) the velocity field decays as $r^{-3}$, while for
pushers/pullers ($\alpha\ne 0$) the velocity field decays as $r^{-2}$.
In contrast, the velocity field for a sedimenting particle (or a
particle experiencing a net body force) decays as
$r^{-1}$\cite{Russel:1992wr}. This will have important consequences
when considering the hydrodynamic interactions of suspensions of
swimmers.

\subsection{Simulation Method : Basic Equations}
\label{s:spm}
We propose a direct numerical simulation (DNS) procedure to study a
system of self-propelled squirmers based on the ``Smoothed Profile''
(SP) method~\cite{Nakayama2005a,Nakayama:2008fi}, which allows one to
efficiently solve both the Navier-Stokes equation (NS), for the fluid
motion, and the Newton-Euler equations, for the colloids. This method
has been successfully used to study the diffusion, sedimentation,
electro-hydrodynamics, and rheology of colloidal dispersions in
incompressible
fluids\cite{Iwashita:2008cj,Hamid:2012ts,Kim2006a,Kobayashi2011a}, and
recent work has shown how it can be extended to treat compressible
fluids within a fluctuating-hydrodynamics
approach\cite{Tatsumi:2012ek}. A detailed error analysis of the method
can be found in~ref.~\citenum{Luo:2009hu}. The basic idea is to replace
the sharp boundary at the colloid/fluid interface with a diffuse
interface of finite thickness $\zeta$. This allows one to discretize
the system using a fixed Cartesian coordinate grid, since the
interface will always be supported by multiple-grid points. Although a
loss of accuracy at the surface of the particle is inevitable, we can
easily impose periodic boundary conditions (PBC), and use a Fourier
spectral method to solve for the fluid equations of motion. The
particles in the SP method are not treated as boundary conditions for the
host fluid, but rather as a body force in the NS equation. Thus, we
avoid the mesh-reconstruction problems that plague most computational
fluid dynamics methods for systems with moving boundaries. We are
aware of two alternative simulation methods that aim to describe these
squirmer suspensions at the same level of description, the first was
developed by Ramachandran et al.\cite{Ramachandran:2006jk} using a
Lattice Boltzmann (LB) model, and the second was originally introduced
by Downton and Stark\cite{Downton:2009ig} within a multi-particle
collision dynamics (MPC) framework, and later extended by G\"{o}tze
and Gompper\cite{Gotze:2010jl} to recover the correct rotational
dynamics. Although the implementation details are
specific to each of the models (LB, MPC, SP), the general mechanism
used to obtain the squirming motion is the same in all three cases:
local conservation of momentum. For the moment though, these DNS
approaches have not been extensively used to study these types of
swimming systems; the most popular methods, which still account for
the hydrodynamic interactions, have usually been based on Stokesian
Dynamics\cite{BRADY:1988up,Swan:2011eq,Ishikawa:2006hf,Ishikawa:2007cf,Ishikawa:2007gk,Ishikawa:2008ht,Ishikawa:2010ho},
and are thus limited to Newtonian fluids in the Stokes regime.

In what follows, we briefly review the governing equations for a
dispersion of inert colloids in a simple Newtonian-fluid, before
considering how the equations must be modified for use with the SP method
for swimming particles with slip boundary conditions. The formulation
we present closely follows that of~ref.~\citenum{Nakayama:2008fi}, in
which a more detailed description of the computational algorithm can
be found. The motion of the host fluid is determined by the
Navier-Stokes equation with the incompressibility condition
\begin{align}
  \nabla\cdot\vec{u}_f &= 0 \label{e:ns_sole}\\
  \rho\left(\partial_t + \vec{u}_f\cdot\nabla\right)\vec{u}_f &=
  \nabla\cdot\tensor{\sigma}\label{e:ns}
\end{align}
where $\rho$ is the total mass density of the fluid,
$\vec{u}_f$ is the host fluid velocity field, and
$\tensor{\sigma}$ is the stress tensor
\begin{align}
  \tensor{\sigma} &= -p\tensor{I} + \tensor{\sigma}^\prime \label{e:stress}\\
  \tensor{\sigma}^\prime &= \eta\left[\nabla\vec{u}_f +
    \left(\nabla\vec{u}_f\right)^{\text{t}}\right]\label{e:visc_stress}
\end{align}
with $\eta$ the shear viscosity of the fluid. Consider a mono-disperse
system of $N$-spherical particles, of radius $a$, mass $M_p$, and
moment of inertia $\tensor{I}_p=2/5 M_pa^2\tensor{I}$ (with
$\tensor{I}$ the unit tensor). The evolution of the colloids is given
by the Newton-Euler equations\cite{Jose:1998vq},
\begin{align}
  \dt{\vec{R}}_i &= \vec{V}_i\,
  &\dt{\tensor{Q}_i} &=
  \tensor{Q}_i\,\text{skew}\left(\tensor{\Omega}_i\right)\label{e:newton}\\
  M_p\dt{\vec{V}_i} &= \vec{F}_i^{\text{H}} + \vec{F}_i^{\text{C}} +
  \vec{F}_i^{\text{ext}} &\tensor{I}_p\cdot\dt{\vec{\Omega}}_i &=
  \vec{N}_i^{\text{H}} + \vec{N}^{\text{ext}}\notag
\end{align}
where $\vec{R}_i$ and $\vec{V}_i$ denote the center of mass position
and velocity of particle $i$, respectively, $\tensor{Q}_i$ is the
orientation matrix\footnote[2]{For numerical stability we use
  quaternions, and not rotation matrices, to represent the rigid body
  dynamics of the particles} and $\vec{\Omega}_i$ the angular
velocity, with $\text{skew}\left(\vec{\Omega}_i\right)$ the
skew-symmetric angular velocity matrix
\begin{equation}\label{e:skew}
  \text{skew}(\vec{\Omega}_i) = \begin{pmatrix}
    0 & -\Omega_i^{z} & \phantom{-}\Omega_i^{y} \\
    \phantom{-}\Omega_i^{z} & 0 & -\Omega_i^{x} \\
    -\Omega_i^{y} & \phantom{-}\Omega_i^{x} & 0
  \end{pmatrix}
\end{equation}
The forces on the particles are comprised of hydrodynamic
contributions arising from the fluid-particle interactions
$\vec{F}^{\text{H}}$, the colloid-colloid interactions due to the core
potential of the particles $\vec{F}^{\text{C}}$ (which prevents
particle overlap), and a possible external field contribution
$\vec{F}^{\text{ext}}$ (such as gravity). Likewise, the torques on the
particles can be divided into a hydrodynamic $\vec{N}^{\text{H}}$ and
an external contribution $\vec{N^{\text{ext}}}$ (for simplicity, the
particle-particle interactions are assumed to be given by a radial
potential). In what follows we consider buoyancy-neutral particles, so
that $\vec{F}^{\text{ext}} = \vec{N}^{\text{ext}} = 0$. Finally,
conservation of momentum between the fluid and the particles implies
the following hydrodynamic force and torque on the $i$-th particle
\begin{align}
  \vec{F}_i^{\text{H}} &=
  \int\df{\vec{S}_i}\cdot\tensor{\sigma} \label{e:forces}\\
  \vec{N}_i^{\text{H}} &= \int\left(\vec{x} -
    \vec{R}_i\right)\times\left(\vdf{S}_i\cdot\tensor{\sigma}\right)
  \label{e:torques}
\end{align}
where $\int\df{\vec{S}_i}$
indicates an integral over the particle surface.

\subsection{Simulation Method: Smoothed Profile Squirmers}
We now present the computational algorithm used to simulate the motion
of spherical particles, with a given surface tangential slip velocity
$\vec{u}^s$, using the SP method. We require that all field variables be
defined over the entire computational domain (fluid + particle). The
concentration field for the colloids is given as $\phi(\vec{x}, t) =
\sum_{i=1}^N \phi_i(\vec{x},t)$, where $\phi_i\in\left[0,1\right]$ is
the smooth profile field of particle $i$. This field is defined such
that it is unity within the particle domain, zero in the fluid domain,
and smoothly interpolates between both within the interface regions.
Details on the specific definition and the properties of this profile
function can be found in~ref.~\citenum{Nakayama2005a}. The particle
velocity field is defined in a similar fashion, as
\begin{equation}\label{e:phi_up}
  \phi\vec{u}_p(\vec{x},t) = \sum_{i=1}^N\left\{
    \vec{V}_i(t) + \vec{\Omega}_i(t)\times\vec{r}_i(t)
  \right\}\phi_i(\vec{x},t)
\end{equation}
with $\vec{r}_i = \vec{x} - \vec{R}_i$, which allows one to define the
total fluid velocity field as
\begin{equation}\label{e:utot}
  \vec{u}(\vec{x},t) \equiv (1-\phi)\vec{u}_f + \phi\vec{u}_p
\end{equation}
where the incompressibility condition is satisfied on the entire domain
$\nabla\cdot\vec{u} = 0$. The evolution equation for $\vec{u}$ is then
derived assuming momentum-conservation between fluid and particles~\cite{Nakayama2005a}
\begin{equation}\label{e:ns_sp}
  \rho\left(\partial_t + \vec{u}\cdot\nabla\right)\vec{u} =
  \nabla\cdot\tensor{\sigma} + \rho\phi\vec{f}_p + \rho\vec{f}_{sq}
\end{equation}
where $\phi\vec{f}_p$ represents the force density field needed to
maintain the rigidity constraint on the particle velocity field and
$\vec{f}_{sq}$ is the force density field generated by the particles'
squirming motion. 

We use a fractional step approach to update the
total velocity field. Let $\vec{u}^n$ be the field at time $t_n=n h$
($h$ is the time interval). We first solve for the advection and
hydrodynamic viscous stress terms, and propagate the particle
positions (orientations) using the current particle velocities. This
yields
\begin{align}
  \vec{u}^* &= \vec{u}^n + \int_{t_n}^{t_n+h}\df{s}\nabla\cdot\left[
    \frac{1}{\rho}\left(-p^*\tensor{I} + \tensor{\sigma}^\prime\right)
    -
    \vec{u}\vec{u}\right]\label{e:ns_update1}\\
  \vec{R}_i^{n+1} &= \vec{R}_i^{n} +
  \int_{t_n}^{t_n+h}\df{s}\vec{V}_i\label{e:ns_update1_r}\\
  \tensor{Q}_i^{n+1} &= \vec{Q}_i^{n} + \int_{t_n}^{t_n+h}
  \df{s}\tensor{Q}_i\text{skew}\left(\vec{\Omega}_i\right)\label{e:ns_update1_q}
\end{align}
where the pressure term $p^*$ in Eq.~\eqref{e:ns_update1} is
determined by the incompressibility condition $\nabla\cdot\vec{u}^* = 0$.
The remaining updating procedure applies to the slip
condition at the particle boundary and the rigidity constraint on the
velocity field.

We now consider the momentum change needed to maintain the slip
velocity at the surface of each of the squirmers, where the slip
profile $\vec{u}^s$ is imposed with respect to the particle velocities
$\{\vec{V}_i^\prime;\vec{\Omega}_i^\prime\}$, using the previously
updated positions and orientations
$\{\vec{R}_i^{n+1};\tensor{Q}_i^{n+1}\}$. We note that at this point
we do not yet know the updated particle velocities
$\{\vec{V}_i^{n+1};\vec{\Omega}_i^{n+1}\}$, which are the values that
should be used when enforcing the surface slip profile
$\vec{V}_i^\prime = \vec{V}_i^{n+1}$ ($\vec{\Omega}_i^\prime =
\vec{\Omega}_i^{n+1}$). Therefore, we adopt an iterative solution, and
as an initial guess, we use the particle velocities at the previous
time step, i.e., $\vec{V}_i^\prime = \vec{V}_i^n$
($\vec{\Omega}_i^\prime = \vec{\Omega}_i^n$). The updated total
velocity field is now
\begin{align}
  \vec{u}^{**} &= \vec{u}^{*} +
  \left[\int_{t_n}^{t_n+h}\df{s}\vec{f}_{\text{sq}}\right] \label{e:ns_update2_b}\\
  \left[\int_{t_n}^{t_n+h}\df{s}\vec{f}_{\text{sq}}\right] &= \vec{u}^{*} +
  \sum_{i=1}^N\varphi_i\left(
    \vec{V_i}^{\prime} + \vec{\Omega}_i^{\prime}\times\vec{r}_i +
    \vec{u}_i^{s} - \vec{u}^*
  \right)
  + \sum_{i=1}^{N}\phi_i\left(\delta\vec{V}_i +
    \delta\vec{\Omega}_i\times\vec{r}_i\right) - \frac{h}{\rho}\nabla{p_{sq}}\label{e:ns_update2_c}
\end{align}
\begin{figure}[ht!]
  \centering
  \includegraphics[width=0.5\textwidth]{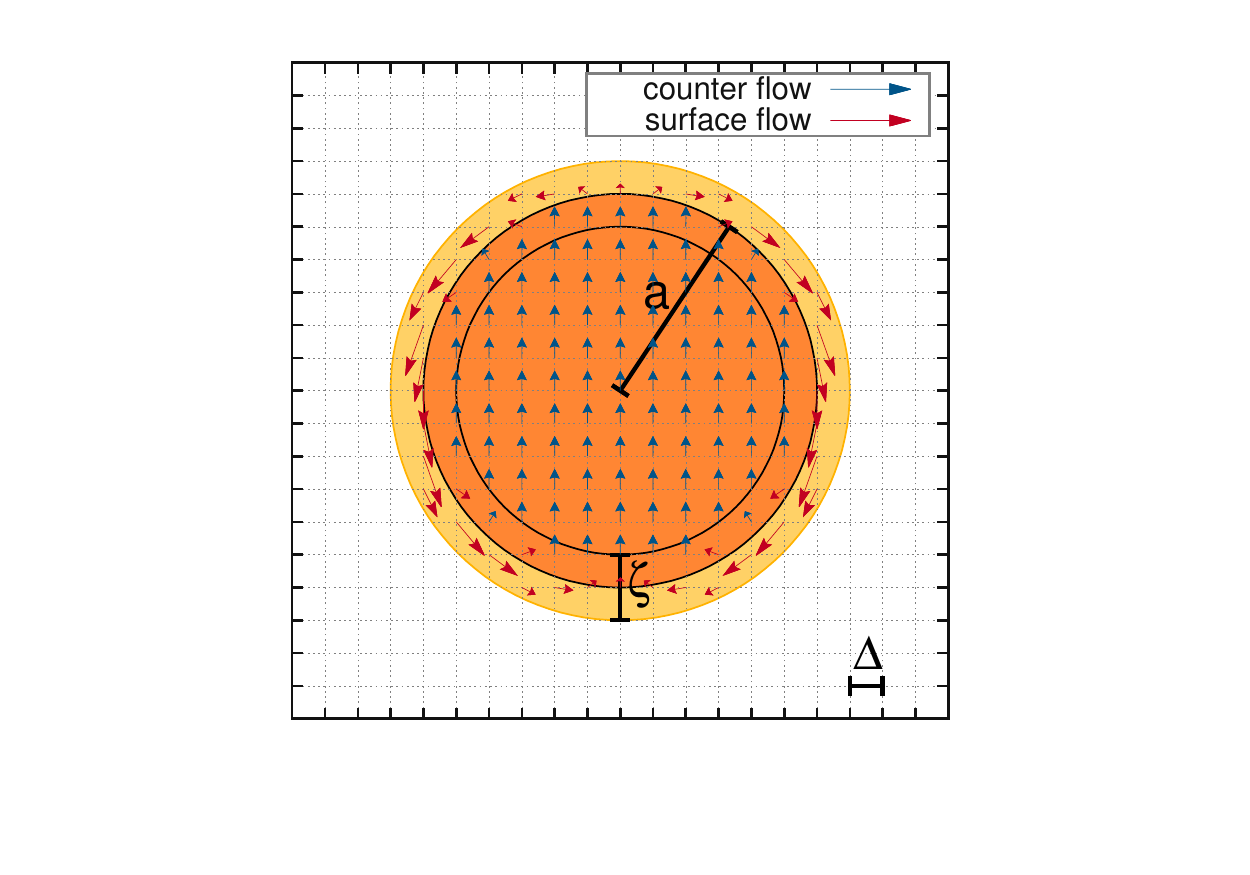}
  \caption{\label{f:slip_vel}(color online) Schematic representation of the updating
    scheme used to enforce the slip boundary condition at the surface
    of the squirmers. Each particle is considered to exert a force on
    the fluid at the surface, in order to maintain the specified flow
    profile $\vec{u}^s$ (red arrows) for the squirming motion. To
    ensure local momentum conservation, a counter-flow is added within the
    particle domain (blue arrows).}
\end{figure}

The second term on the right hand side of Eq.~\eqref{e:ns_update2_c}
imposes the slip velocity profile $\vec{u}^s$ at the surface of each
of the squirmers; where $\varphi_i \propto (1-\phi_i) \norm{\nabla
  \phi_i}$ is a smooth surface profile function which is non-zero only
within the interface domain of the squirmer (normalized to have a
maximum value of one), and zero everywhere else. The third term adds a
counter-flow entirely within the particle domain, in
such a way that local momentum conservation is obtained. Assuming
rigid-body motion, with velocities $\delta\vec{V}_i$ and
$\delta\vec{\Omega}_i$, this requires 
\begin{align}
  \int\vdf{x}\phi_i\left(\delta\vec{V}_i +
    \delta\vec{\Omega}_i\times\vec{r}_i\right) &= -
  \int\vdf{x}\varphi_i\left(\vec{V}_i^{\prime} +
    \vec{\Omega}_i^{\prime}\times\vec{r}_i + 
    \vec{u}_i^s - \vec{u}^*\right) \label{e:droplet}\\
  \int\vdf{x}\vec{r}_i\times\phi_i\left(\delta\vec{V}_i +
    \delta\vec{\Omega}_i\times\vec{r}_i\right) &= -
  \int\vdf{x}\vec{r}_i\times \varphi_i\left(\vec{V}_i^{\prime} +
    \vec{\Omega}_i^{\prime}\times\vec{r}_i + \vec{u}_i^s -
    \vec{u}^*\right)\label{e:droplet_torque}
\end{align}
from which we can easily obtain the counter-flow terms
$\delta\vec{V}_i$ ($\delta\vec{\Omega}_i$) from the particle
velocities $\vec{V}_i^\prime$ ($\vec{\Omega}_i^\prime$). A schematic
representation of this procedure, used to enforce the specific
slip-boundary conditions for our model squirmers, is shown in
Figure~\ref{f:slip_vel}. Finally, the pressure term due to the
squirming motion $p_{sq}$ is obtained from the incompressibility
condition $\nabla\cdot\vec{u}^{**} = 0$. At this point, the momentum
conservation is solved for the total velocity field.

The hydrodynamic force and torque exerted by the fluid on the colloids
(which includes all contributions due to the squirming motion) is
again derived by assuming momentum conservation. The time integrated
hydrodynamic force and torque over a period $h$ are equal to the
momentum exchange over the particle domain
\begin{align}
  \left[\int_{t_n}^{t_n+h}\df{s}\left(\vec{F}_i^{\text{H}} +
      \vec{F}_i^{\text{sq}}\right)\right] &=
  \int\vdf{x}\rho\phi_i^{n+1}\left(\vec{u}^{**} - \vec{u}_p^{n}\right)\label{e:momentum}
  \\
  \left[\int_{t_n}^{t_n+h}\df{s}\left(\vec{N}_{i}^{\text{H}} +
      \vec{N}_i^{\text{sq}}\right)\right] &=
  \int\vdf{x}\left[\vec{r}_i^{n+1}\times\rho\phi_i^{n+1}\left(\vec{u}^{**}-\vec{u}_p^{n}\right)\right]\label{e:ang_momentum}
\end{align}
From this and any other forces on the colloids, the particles
velocities are updated as
\begin{align}
  \vec{V}_i^{n+1} &= \vec{V}_i^{n} +
  M_p^{-1}\left[\int_{t_n}^{t_n+h}\df{s}\left(\vec{F}_i^{\text{H}} +
      \vec{F}_i^{\text{sq}}\right)\right] +
  M_p^{-1}\left[\int_{t_n}^{t_n+h}\df{s}
    \left(\vec{F}_i^{\text{C}}+\vec{F}_i^{\text{ext}}\right)\right]\label{e:vp_upd}\\
  \vec{\Omega}_i^{n+1} &= \vec{\Omega}_i^{n} + \tensor{I}_p^{-1}\cdot
  \left[\int_{t_n}^{t_n+h}\df{s}\left(\vec{N}_i^{\text{H}} +
      \vec{N}_i^{\text{sq}}\right)\right]
  +\tensor{I}_p^{-1}\cdot\left[\int_{t_n}^{t_n+h}\df{s}\vec{N}_i^{\text{ext}}\right]\label{e:wp_upd}
\end{align}
We recall that we have imposed the slip profile $\vec{u}^{s}$ with
respect to the primed velocities
$\{\vec{V}_i^\prime;\vec{\Omega}_i^\prime\}$, which need not be equal
to the final velocities of the particle at step $n+1$. To maintain
consistency, we iterate over
Eqs.~\eqref{e:ns_update2_b}-\eqref{e:wp_upd} until convergence in the
velocities is achieved. Finally, the resulting particle velocity field
$\phi^{n+1}\vec{u}_p^{n+1}$ is enforced on the total velocity field as
\begin{align}
  \vec{u}^{n+1} &= \vec{u}^{**} +
  \left[\int_{t_n}^{t_n+h}\df{s}\phi\vec{f}_p\right]\label{e:rigidity}\\
  \left[\int_{t_n}^{t_n+h}\df{s}\phi\vec{f}_p\right] &=
  \phi^{n+1}\left(\vec{u}_p^{n+1}-\vec{u}^{**}\right) -
  \frac{h}{\rho}\nabla{p_{\text{p}}}\label{e:rigidity_frc}
\end{align}
with the pressure due to the rigidity constraint obtained from the
incompressibility condition $\nabla\cdot\vec{u}^{n+1} = 0$. The total
pressure field is then given by $p = p^* + p_{\text{p}} +
p_{\text{sq}}$.
\subsection{Choosing the Relevant Reference Frame}
Although we have not included thermal fluctuations in our system, the
swimming motion of the particles is known to give rise to diffusive
behavior at sufficiently high particle
concentrations\cite{Ishikawa:2007cf}. However, due to the
self-propelled nature of the particles, the largest contribution to
their displacement will naturally come from the their inherent
swimming. As such, one must wait a very long time before the particles
exhibit any type of diffusive behavior, and even then, it is difficult
to establish what role the hydrodynamic interactions among neighboring
particles are playing. All particles will be swimming in the flow
field generated by their neighbors, and the interactions among them
can give rise to motion perpendicular to the particle's preferred
swimming direction, as well as provide a momentary impulse that can
increase/decrease the velocity parallel to the swimming axis, or even
change its orientation in space. In order to better understand this
phenomena, we analyze the particle motion with respect to the frame of
reference of the moving squirmers, as was proposed by Han et
al.\cite{Han:2006ew} to study the Brownian motion of ellipsoidal
particles. We begin by decomposing the particle trajectories in terms
of displacements between successive time intervals, which we take to
be equally spaced. If $\vec{R}(t_n)$ denotes the position of a tagged
particle at time step $n$ ($t_n = n \delta t$, with $\delta t$ a
suitably small time interval), we can express the time evolution of
its position $\{\vec{R}(t_i)\}$ as
\begin{equation}
  \{\vec{R}(t_0), \vec{R}(t_0)+\Delta_1, \ldots, \vec{R}(t_0) +
  \sum_{j=1}^{i}\Delta_j, \ldots, \vec{R}(t_0) + \sum_{j=1}^{n}\Delta_n\}
\end{equation}
where $\Delta_i = \vec{R}(t_i) - \vec{R}(r_{i-1})$. Using this
notation, the translational diffusivity is given as\cite{Hansen}
\begin{align}
  D_T(t_n) &= \frac{1}{6 n \delta t} \avg{\left(\vec{R}(t_n) -
      \vec{R}(t_0)\right)^2}\label{e:diff}
  \\
  &= \frac{1}{6 n \delta t}
  \avg{\left[\sum_{i=1}^n\Delta_n\right]^2}\notag
\end{align}
To study the coupling between translation and rotation of the
particles, we will also consider the rotational diffusivity, defined
as
\begin{align}
  D_R(t_n) &= \frac{1}{6 n \delta
    t}\avg{\left(\vec{\vartheta}{(t_n)}-\vec{\vartheta}{(t_0)}\right)^2} 
  \label{e:rot_diff}\\
  \vec{\vartheta}(t_n) &= \int_{t_0}^{t_n}\df{s} \vec{\Omega}(s)
\end{align}
where $\vec{\vartheta}$ is the unbounded rotational displacement of the
particle.

\begin{figure}[ht!]
  \centering
  \includegraphics[width=0.4\textwidth]{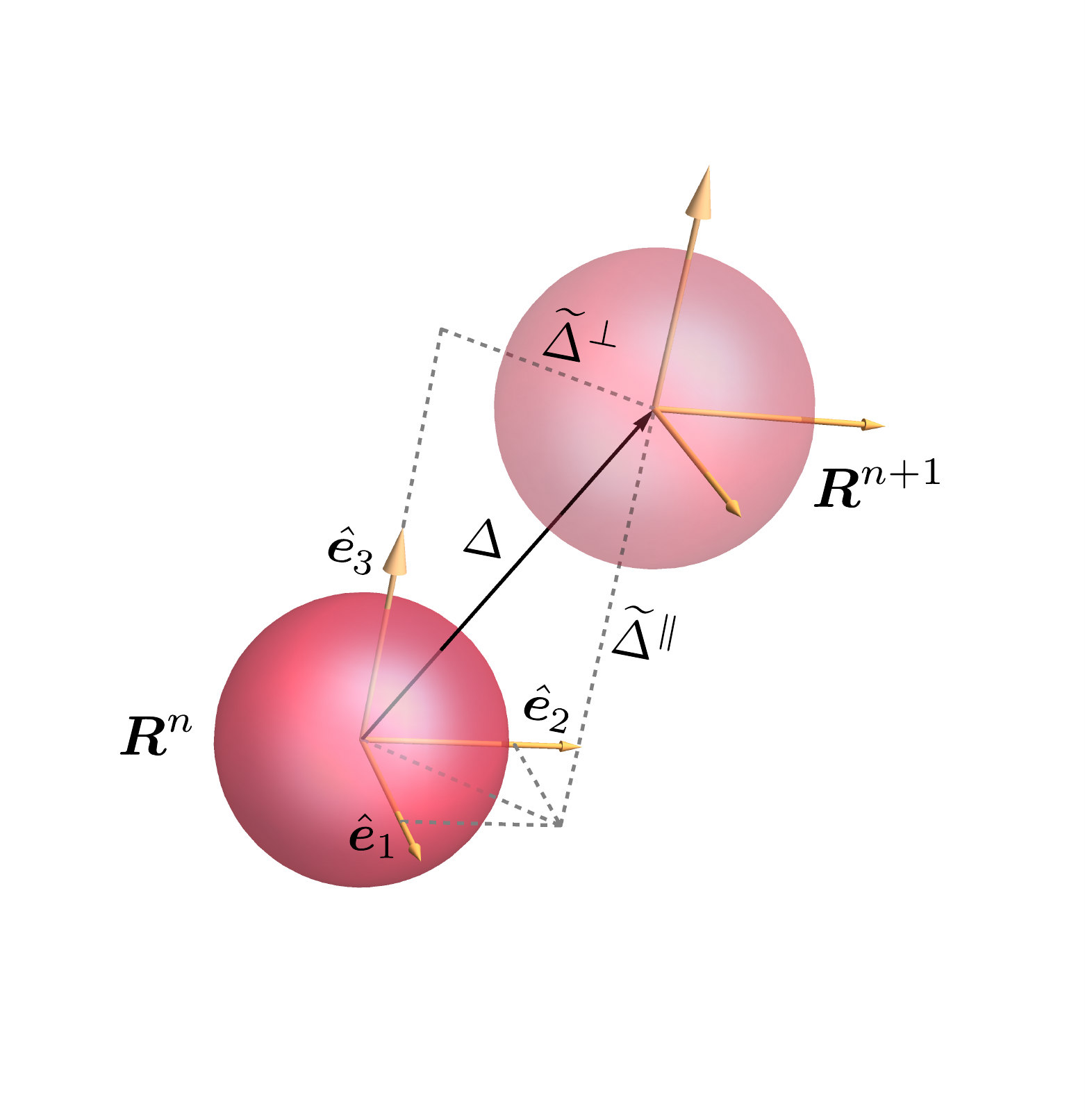}
  \caption{\label{f:body_disp}Analysis of the particle displacements
    with respect to the particle's moving coordinate frame. The
    particle displacement $\Delta$ during a given time interval is
    decomposed into its components parallel $\body{\Delta}^\parallel$
    and perpendicular $\body{\Delta}^\perp$ to the swimming axis. For
    a single isolated squirmer $\body{\Delta}^\perp = 0$ for any given
    time interval; for a suspension of swimmers the flow induced by
    the neighboring particles gives rise to non-zero perpendicular
    displacements.}
\end{figure}
\begin{figure}[ht!]
  \centering
  \includegraphics[width=0.32\textwidth]{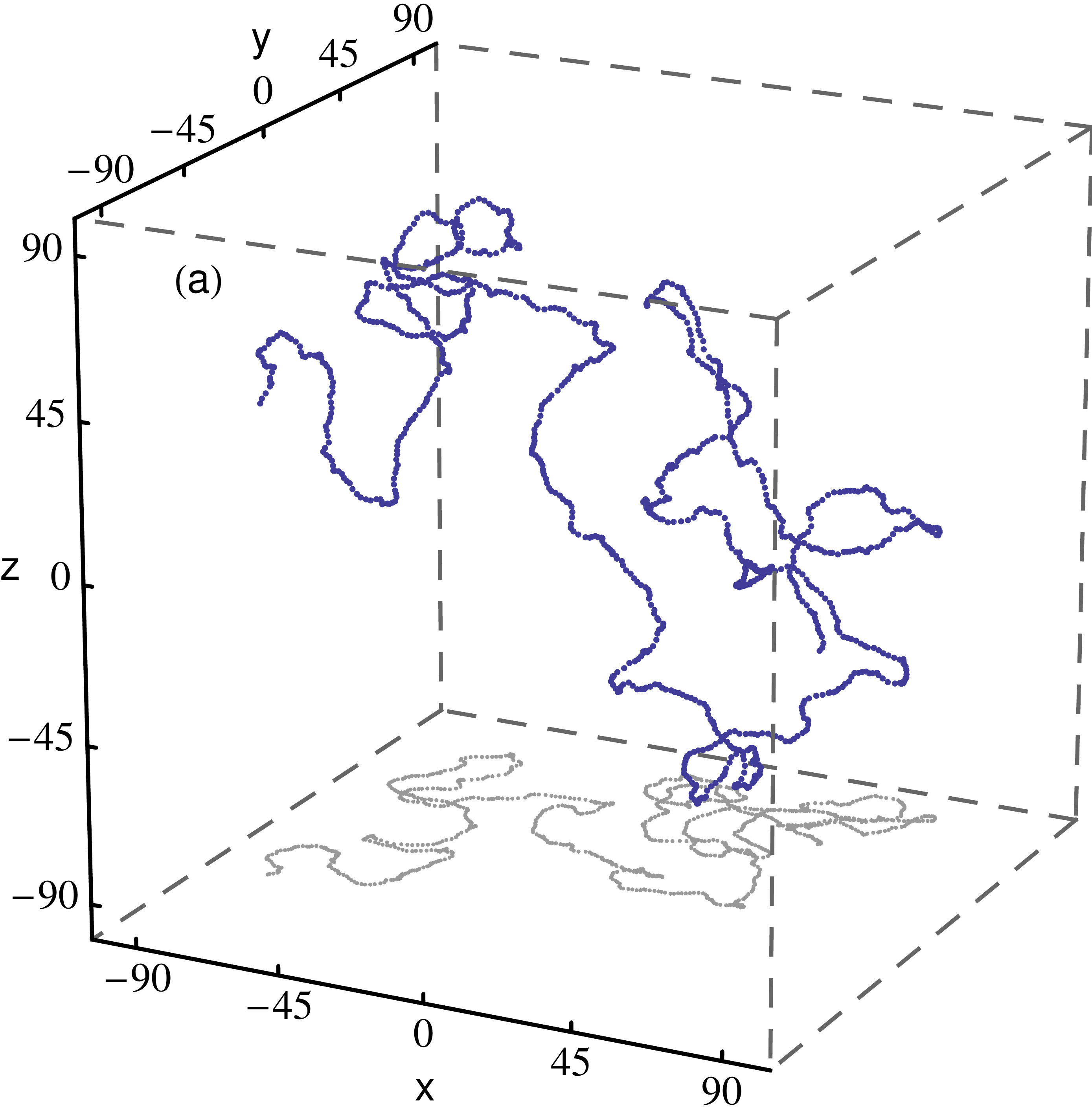}\\
  \includegraphics[width=0.32\textwidth]{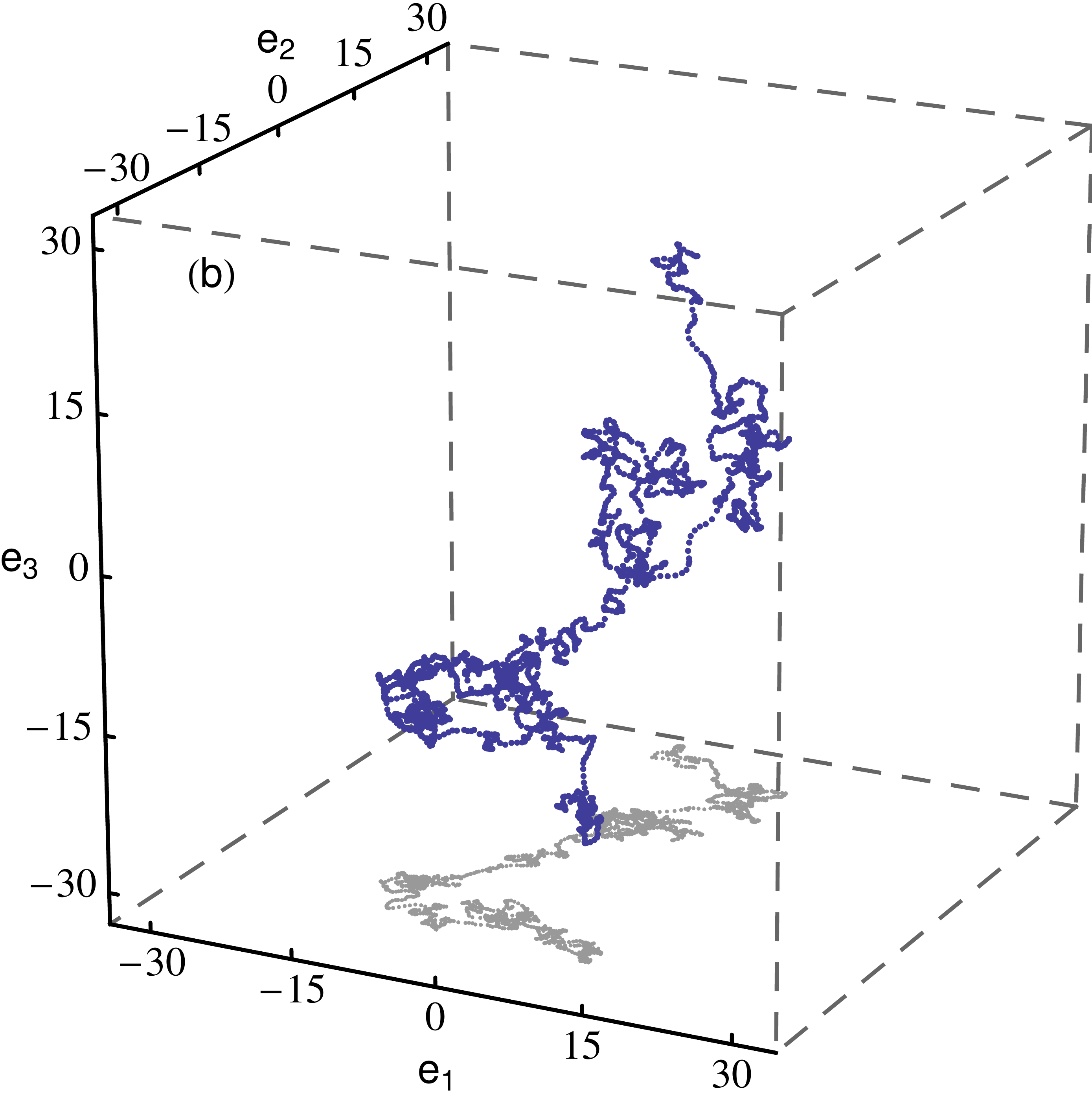}
  \caption{\label{f:body_trj}(color online) Trajectory of a single
    particle in a suspension of pullers $\alpha=+2$ at $\phi=0.1$. (a)
    The \textit{real} lab-space trajectory of the particle, and (b)
    the body-space trajectory. The latter is constructed by
    transforming the individual particle displacements to the frame of
    reference of the particle, and expressing them in terms of
    displacements parallel and perpendicular to the particle's
    swimming axis $\uvec{e}_3$. In addition, the (parallel) motion due
    to the average swimming speed, $\Delta x(t) =
    \avg{\vec{V}\cdot\uvec{e}_3} t$, has been removed. Thus,
    Figure~(b) gives the random motion induced on the particle by the
    local fluctuations in the surrounding flow field. The length scale
    is given by the particle radius $a/\Delta=5$ and the projections
    of the trajectories onto the bottom plane have been added as a
    visual guide.}
\end{figure}

In order to separate the ``random'' contribution given by the
surrounding configuration from the particle's own swimming motion, we
consider the displacements within the body frame of reference,
\begin{equation}
  \body{\Delta}_i = Q_{i-1}^t \Delta_i
\end{equation}
where tildes are used to denote quantities with respect to the
coordinate-system attached to each of the squirmers $(\uvec{e}_1,
\uvec{e}_2, \uvec{e}_3)$, with the $\uvec{e}_3$ the preferential
swimming axis (see Figure~\ref{f:body_disp}). The advantage of this
approach is illustrated in Figure~\ref{f:body_trj}, where the
trajectory of a single particle, from a suspension of pushers
$\alpha=+2$ at $\phi=0.1$, is given in both representations: with
respect to the lab- and body-space displacements. One can immediately
see where the difficulty in analyzing the particle motion comes from,
as the length scales for the directed (swimming) and fluctuating
motion differ by an order of magnitude. We define an additional
\textit{effective} hydrodynamic diffusivity $\body{D}_T$, in terms of these
perpendicular and parallel displacements, as
\begin{align}
  \body{D}_T^{\perp}(t_n) &= \frac{1}{4 n \delta t} \avg{\left[
      \sum_{i=1}^n \body{\Delta}_i^\perp
    \right]^2}\label{e:diff_perp}\\
  \body{D}_T^{\parallel}(t_n) &= \frac{1}{2 n \delta t} \avg{\left[
      \sum_{i=1}^n
      \body{\Delta}_i^{\parallel} - n\delta t\,U\right]^2} \label{e:diff_parallel}\\
  \body{D}_T(t_n) &= \frac{1}{3}\left(2 \body{D}_T^\parallel(t_n) +
    \body{D}_T^\perp(t_n)\right)\label{e:diff_body}
\end{align}
where the second term inside square brackets in
Eq.\eqref{e:diff_parallel} is required to remove the (average)
contribution to the particle displacements from the inherent swimming
motion ($U = \avg{\vec{V}\cdot\uvec{e}_3}$ is the average velocity
along the particle's swimming axis). These diffusivities provide a
direct measure of the strength of the hydrodynamic interactions among
the squirmers. The corresponding diffusion coefficients can be
obtained from the long-time limit (assuming a
plateau has been reached and the limit exists)
\begin{align}
  D_T &= \lim_{t\to\infty} D_T(t) \label{e:DDT}\\
  D_R &= \lim_{t\to\infty} D_R(t) \label{e:DDR}\\
  \body{D}_T &= \lim_{t\to\infty} \body{D}_T(t) \label{e:EDT}
\end{align}
Since the parallel and perpendicular effective diffusion coefficients,
$\body{D}_T^\parallel$ and $\body{D}_T^\perp$, exhibit no deviation
from isotropic behavior (although this could be expected to change if
persistent long-range order appears), we only consider the average
effective diffusion coefficient $\body{D}_T$. A similar analysis can
be performed for the rotational diffusivities, in order to study the
rotations about the three particle axes independently, but it provides
no additional information and will not be presented here.

Finally, we also consider the velocity auto-correlation
functions\cite{Hansen}
\begin{align}
  C_V(t) &=
  \avg{\vec{V}(t)\cdot\vec{V}(t_0)} \label{e:cvt}\\
  C_\Omega(t) &=
  \avg{\vec{\Omega}(t)\cdot\vec{\Omega}(t_0)} \label{e:cwt}
\end{align}
and in particular, the correlation functions for the velocity
components parallel and perpendicular to the swimming axis
\begin{align}
  C_V^\parallel(t) &=
  \avg{\vec{V}^\parallel(t)\cdot\vec{V}^\parallel(t_0)} \label{e:cvt_zz}\\
  \body{C}_V^\parallel(t) &=
  \avg{\vbody{V}^\parallel(t)\cdot \vbody{V}^\parallel(t_0)} \label{e:cvt_ZZ}\\
  \body{C}_V^\perp(t) &=
  \avg{\vbody{V}^\perp(t)\cdot\vbody{V}^\perp(t_0)}\label{e:cvt_XY}
\end{align}
Although Eqs.~\eqref{e:cvt_zz}~and~\eqref{e:cvt_ZZ} both measure the
correlations of the parallel velocity components, the former does so
within the fixed lab frame, while the latter uses the moving body
frame. Therefore, the first will be sensitive to changes in the
magnitude and direction of the velocity vector, while the second will
only register changes in the magnitude. At short times, we can expect
the decay in correlations of the parallel velocity components to be
determined by the local configuration of the suspension (where the
flow generated by the nearby particles can act to enhance or suppress
the swimming motion), and at long times by the particle collisions,
which will reorient the particle's swimming direction. This means the
initial decay should be the same for both $C_V^\parallel$ and
$\body{C}_V^\parallel$, since the particle has not had time to
reorient itself; at long times however, correlations measured within
the lab frame of reference should decay to zero, whereas those
measured within the body reference should reach a finite value
(determined by the average swimming speed of the squirmers). For the decay
in correlations of the perpendicular velocity components
$\body{C}_V^\perp$ we expect a behavior analogous to the
short-time decay of the parallel components, since it is due entirely
to the flow field generated by the neighboring squirmers.

\section{Validation}
\label{s:test}
\begin{figure}[ht!]
  \centering
  \includegraphics[width=0.45\textwidth]{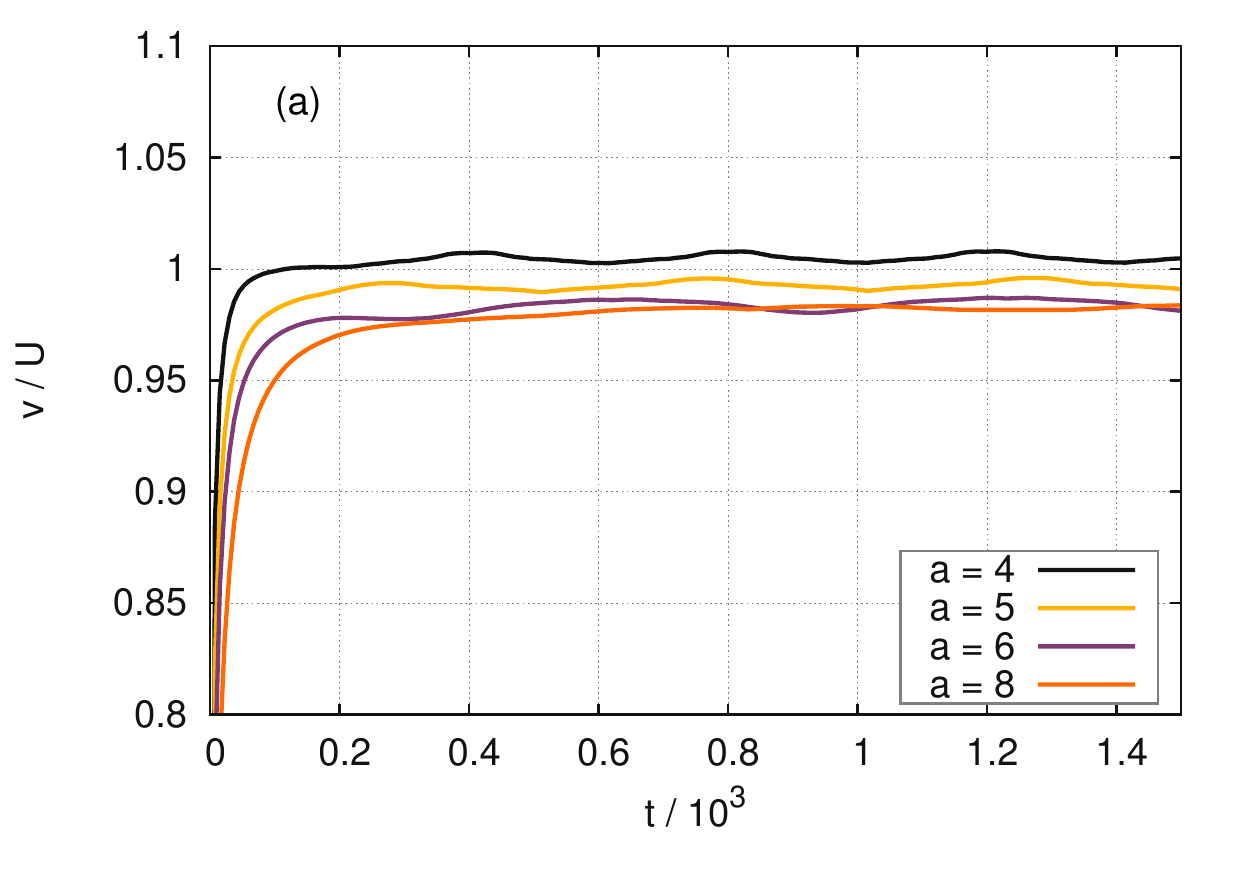}\\
  \includegraphics[width=0.45\textwidth]{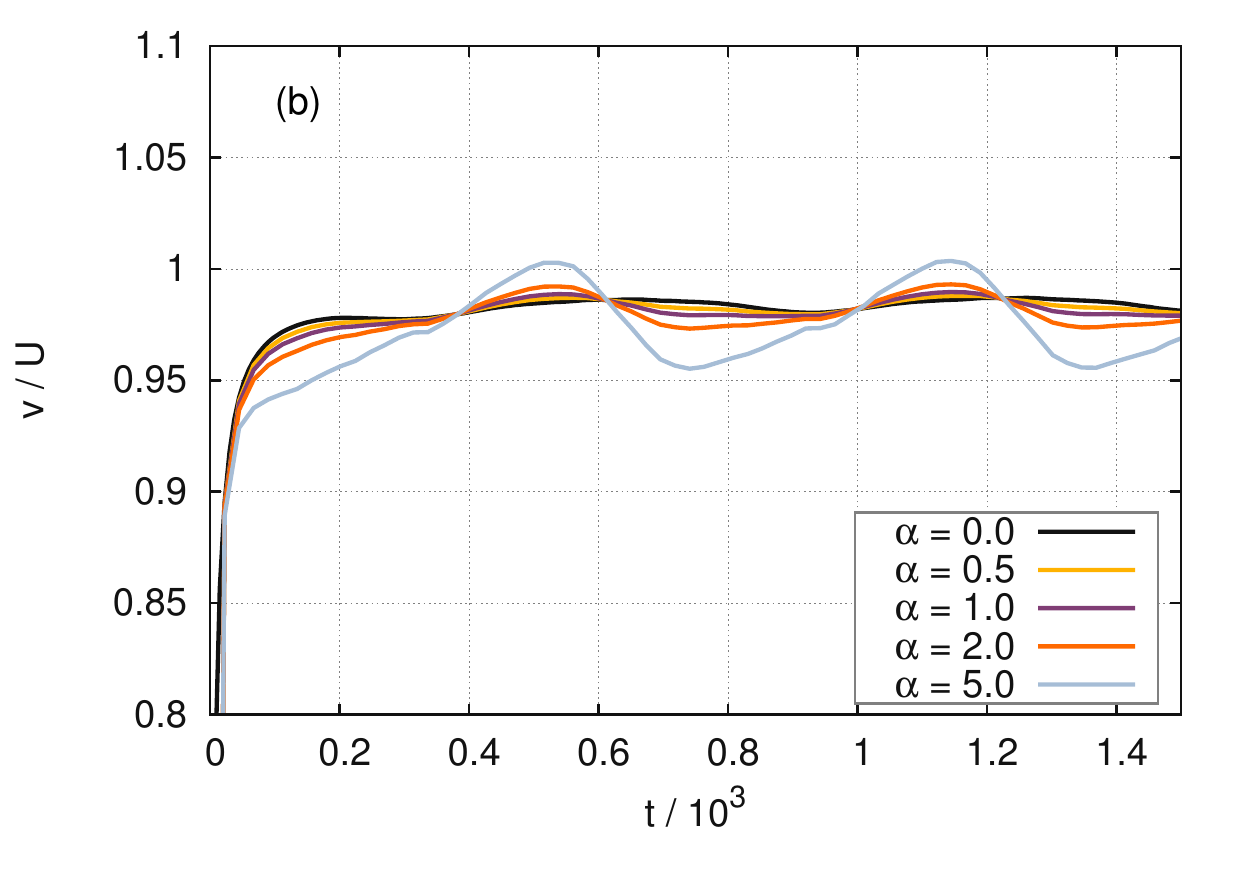}
  \caption{\label{f:swim_speed}(color online) Swimming speed of an
    isolated squirmer in a periodically replicated cubic simulation
    box of length $L/\Delta = 128$ at $\text{Re}=0.01$ as a function
    of time (in simulation units). (a) Neutral swimmer at various
    particle sizes $a/\Delta$. (b) puller of size $a/\Delta = 6$ for
    various swimming modes $0\le \alpha\le 5$. Velocities are scaled
    by the theoretical value for the swimming speed $U=2/3 B_1$.}
\end{figure}
The first obvious test of our simulation method is to make sure that
an isolated swimmer will move at the expected velocity $U=2/3 B_1$,
regardless of the particle size or the value of the second squirming
mode $B_2$. We performed simulations for a single squirmer, inside a
periodically replicated cubic simulation box of dimension
$L=128\Delta$ ($\Delta$ is the grid spacing), for various particles
sizes and squirming modes. Figure~\ref{f:swim_speed} shows the results
obtained for a neutral squirmer ($\alpha=0$), for particle sizes
$a/\Delta = 4, 5, 6, 8$, and for various pullers ($\alpha > 0$), with
a particle size of $a/\Delta = 6$. The particle Reynolds number
$\text{Re}=\rho\, U a / \eta$ and the width of the diffuse interface
$\zeta/\Delta$ were the same for all the simulations, $0.01$ and $2$,
respectively. In all cases, the particle's (average) swimming velocity
shows excellent agreement with the theoretical predictions. For the
neutral squirmers, the swimming velocity is within $\simeq 2\%$ of the
exact value, regardless of the particle diameter, although the speed
shows a small decrease with increasing particle size. The small
oscillations exhibited by the velocity are due to discretization
errors, as the number of grid points on the two hemispheres of the
particle surface will vary depending on the relative position of the
particle center within the computational bins. Similar agreement is
obtained for the pullers (pushers), although the discretization error
increases with increasing $\lvert\alpha\lvert$ (at fixed particle size
$a/\Delta$). As such, for the system sizes we have considered, we are
limited to moderate values of $\lvert\alpha\lvert\lesssim 5$.
\begin{figure}[ht!]
  \centering
  \includegraphics[width=0.33\textwidth]{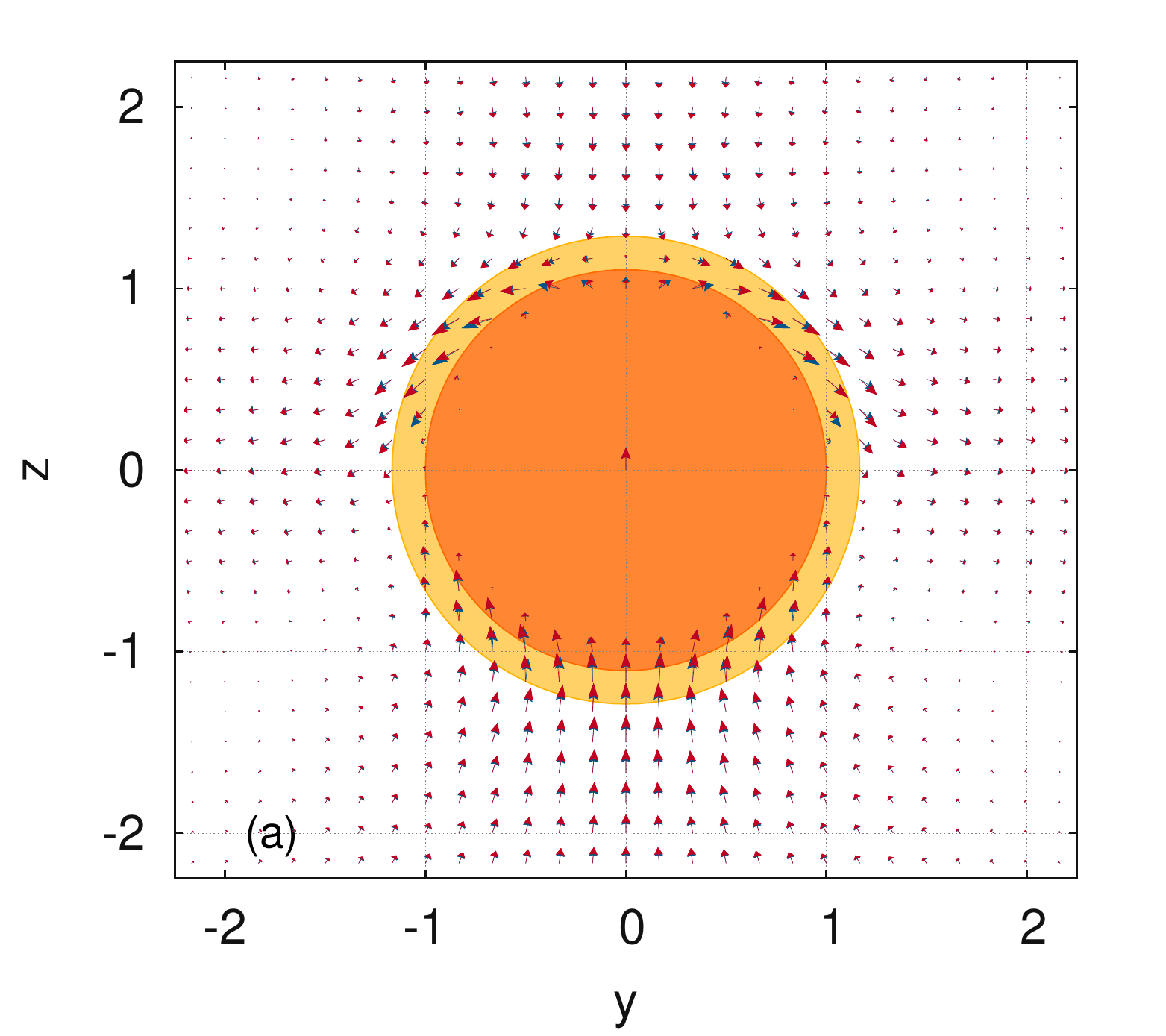}\\
  \includegraphics[width=0.32\textwidth]{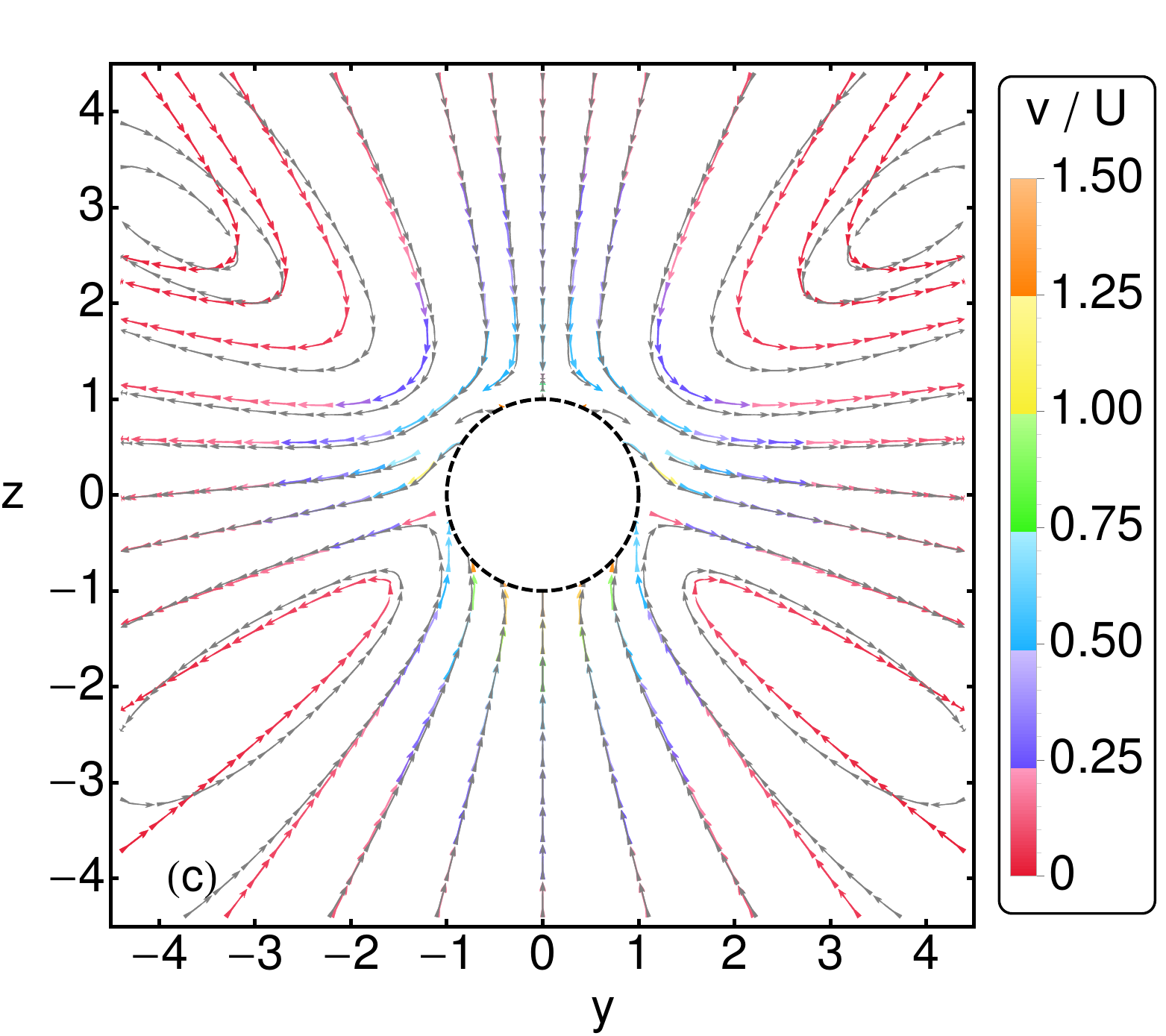}\\
  \includegraphics[width=0.32\textwidth]{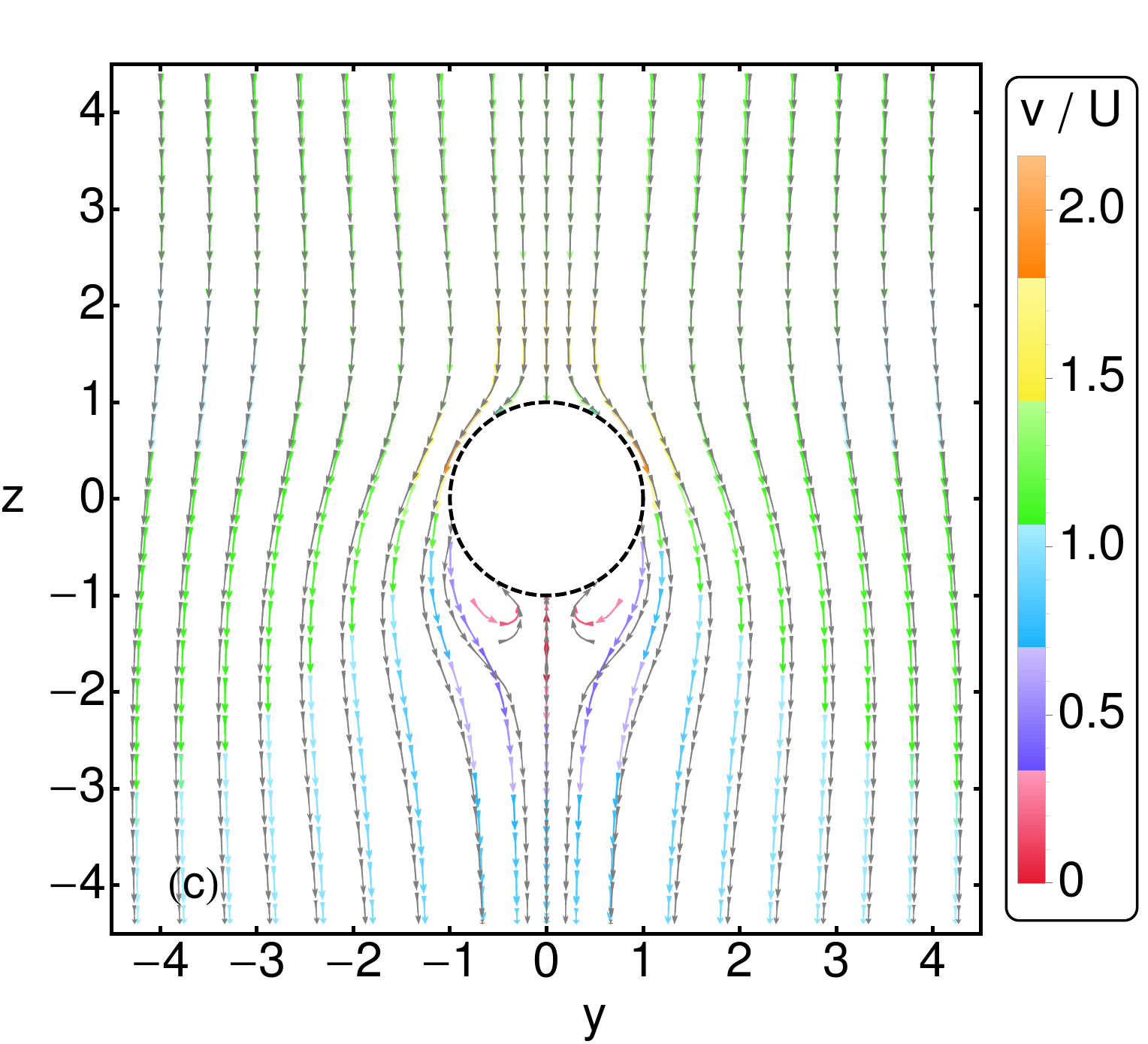}
  \caption{\label{f:stream}(color online) Azymuthaly averaged
    steady-state fluid velocities for a single puller $\alpha=+2$ of
    size $a/\Delta=6$ swimming along the $\uvec{z}$-axis, within a
    periodically replicated cubic simulation box of size
    $L/\Delta=128$. (a) Fluid velocity vectors within the laboratory
    frame, the red and blue arrows show the simulation and analytical
    results, respectively. (b) Fluid velocity stream lines within the
    laboratory frame, the colored lines (color-coded with
    respect to the magnitude of the velocity) represent the analytical
    solution, while the gray lines show the simulation results. (c)
    Same as Figure~(b) but within a reference frame moving with the
    particle. Due to discretization errors, streamlines can begin/end
    at the fluid-particle interface. Length scales have been scaled by
    the particle radius.}
\end{figure}

To verify that the velocity field generated by our SP squirmers is
correct, we compare with the exact (at $\text{Re}=0$) analytical
expression given in Eq.~\eqref{e:squirm_field}. The steady-state
velocity fields generated by a single squirmer ($L/\Delta=128$,
$A/\Delta=6$, $\zeta/\Delta=2$, $\alpha=+2$, $\text{Re}=0.01$), along
with the corresponding stream line plot (in the lab and particle
reference frames), are shown in Figure~\ref{f:stream}. Excellent
agreement with the analytical results is obtained, although
differences in the stream lines arise at large distances $r/a \simeq
4$ for $\theta\simeq\pm \pi/4$ (with respect to the swimming axis
$+\uvec{z}$). Even though the fluid velocity within these regions is
vanishingly small (compared to the swimming speed of the squirmer), a
clear systematic deviation is observed in the direction of the stream
lines. This is due to our use of periodic boundary conditions, which
causes the stream lines to close in on themselves, in order to match at
the boundaries of the simulation cell. A similar deviation is observed
for the velocities along the swimming direction, but these occur at
much larger distances.

\begin{figure}[ht!]
  \centering
  \includegraphics[width=0.45\textwidth]{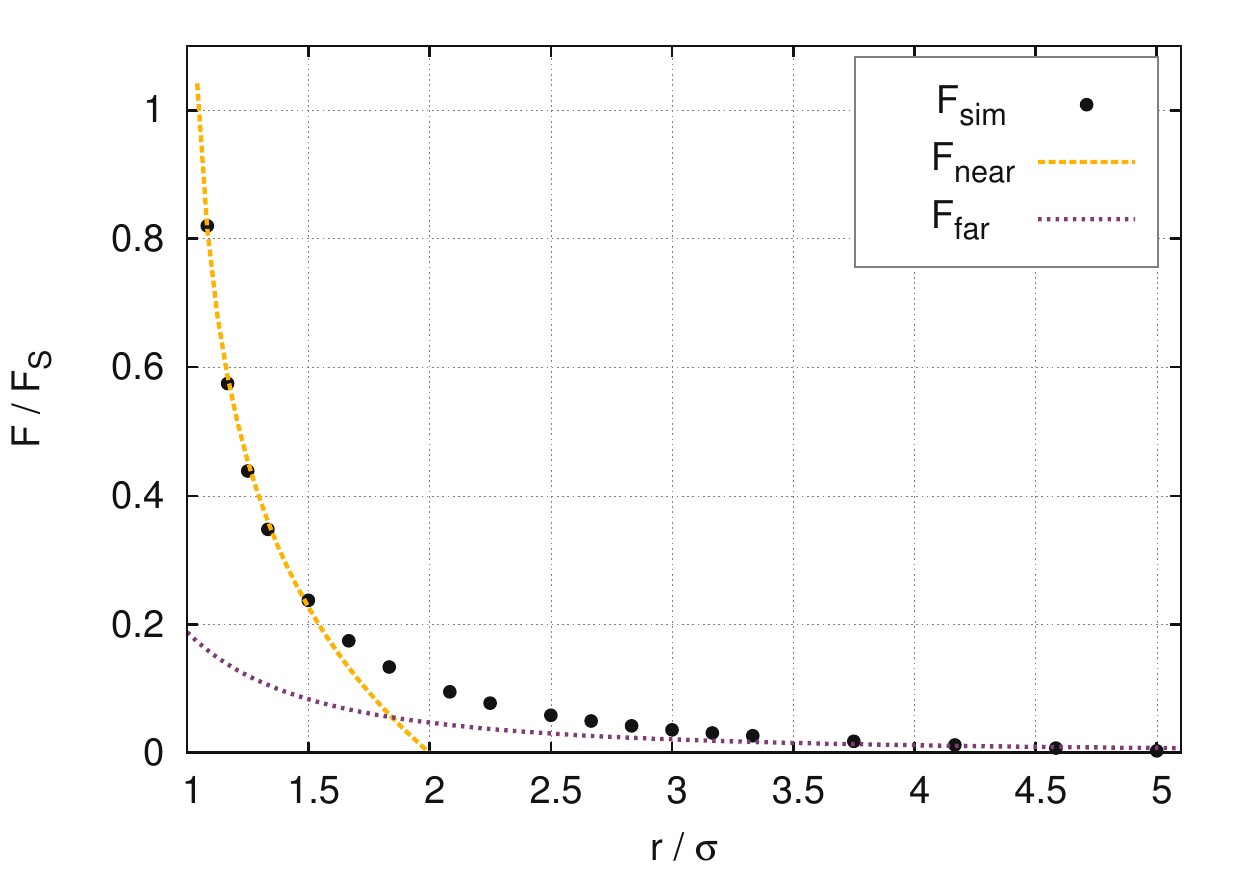}
  \caption{\label{f:pair_force}(color online) Perpendicular force felt
    by two (fixed) parallel squirmers ($\alpha=+2$), of diameter
    $\sigma/\Delta = 10$ and interface thickness $\zeta/\Delta=2$, as
    a function of distance $r$, for $\text{Re}=0.01$ and a system size
    of $L/\Delta=128$. The dashed yellow line gives the fit to the
    expected functional form $F/F_s = A\ln(r/\sigma - 1)$ of the
    near-field force, with $A=-0.327$, while the dotted violet line
    gives the (exact) far-field force. All forces are given in units
    of the Stokes force for an inert sphere with the same speed
    ($U=2/3 B_1$).}
\end{figure}
As in Ref.~\citenum{Gotze:2010jl}, we consider the
interactions between two fixed squirmers at a distance $r$ from each
other, with parallel orientations. Figure~\ref{f:pair_force} shows the
results we have obtained for the force parallel to the displacement
vector between two pullers (perpendicular to their swimming axes),
normalized by the Stokes force $F_s = 6\pi\eta a U$ for an inert
particle moving with the same velocity ($U = 2/3 B_1$). The
simulations were carried out at $\text{Re}=0.01$ for a box size of
$L/\Delta=128$, with a particle radius of $a/\Delta = 6$, and a swimming
mode $\alpha = +2$. The functional form for this perpendicular force
has been given by Ishikawa et al.\cite{Ishikawa:2006hf}
\begin{equation}\label{e:flr}
  F_{\text{near}} \propto \log{\left(\epsilon\right)}
\end{equation}
where $\epsilon = r - \sigma$ is the minimum separation distance
between the surface of the particles (with $\sigma=2 a$ the particle
diameter). Additionally, they also obtained exact far-field
expressions for the force between the two squirmers from a
generalization of Faxen's laws,
\begin{equation}
  F_{\text{far}} = \frac{3\pi\sigma^3 \alpha B_1}{16 r^2}
\end{equation}
At short distances the repulsive force
experienced by the two squirmers is seen to follow the expected
scaling relationship up to $r\lesssim 1.5\sigma$, while the far-field
force is approached asymptotically for $r\gtrsim 3\sigma$. The force
is observed to be proportional to the swimming mode, such that
$F\propto \alpha$ and $\vec{F}(\alpha) = -\vec{F}(-\alpha)$. 
\section{Results and Discussions}
\label{s:main}
\subsection{Diffusion Coefficients}
We have studied the diffusive behavior of a semi-dilute suspension of
identical non-buoyant squirmers, for various concentrations $\phi\le
0.124$ (with $\phi=4\pi a^3 N /3 V$ the packing fraction) and
squirming modes $\alpha=0,\pm 1, \pm 2$. We work in reduced units, in
which the density and viscosity of the host fluid are unity
$\rho=\mu=1$. All simulations were performed for squirmers of radius
$a/\Delta = 5$ (interface width $\zeta/\Delta = 2$), in a cubic
simulation box of length $L/\Delta = 64$, for a particle Reynolds
number of $\text{Re} = \rho\, U a / \mu = 0.05$ ($U$ is the swimming
speed of an isolated squirmer). Although our systems are not very
large $L/a \sim 12$, finite size effects have been shown to be
small\cite{Ishikawa:2007cf}, and since we will be mainly focused on
studying the short-range hydrodynamic interactions among
particles, they can be safely ignored. In what follows, all quantities
are presented in non-dimensionalized form, using as characteristic
length, speed, and time units the particle radius $a$, swimming speed
$U=2/3 B_1$, and the time required for an isolated squirmer to move a
distance equal to its radius $T=a/U$. 
\begin{figure}[ht!]
  \centering
  \includegraphics[width=0.45\textwidth]{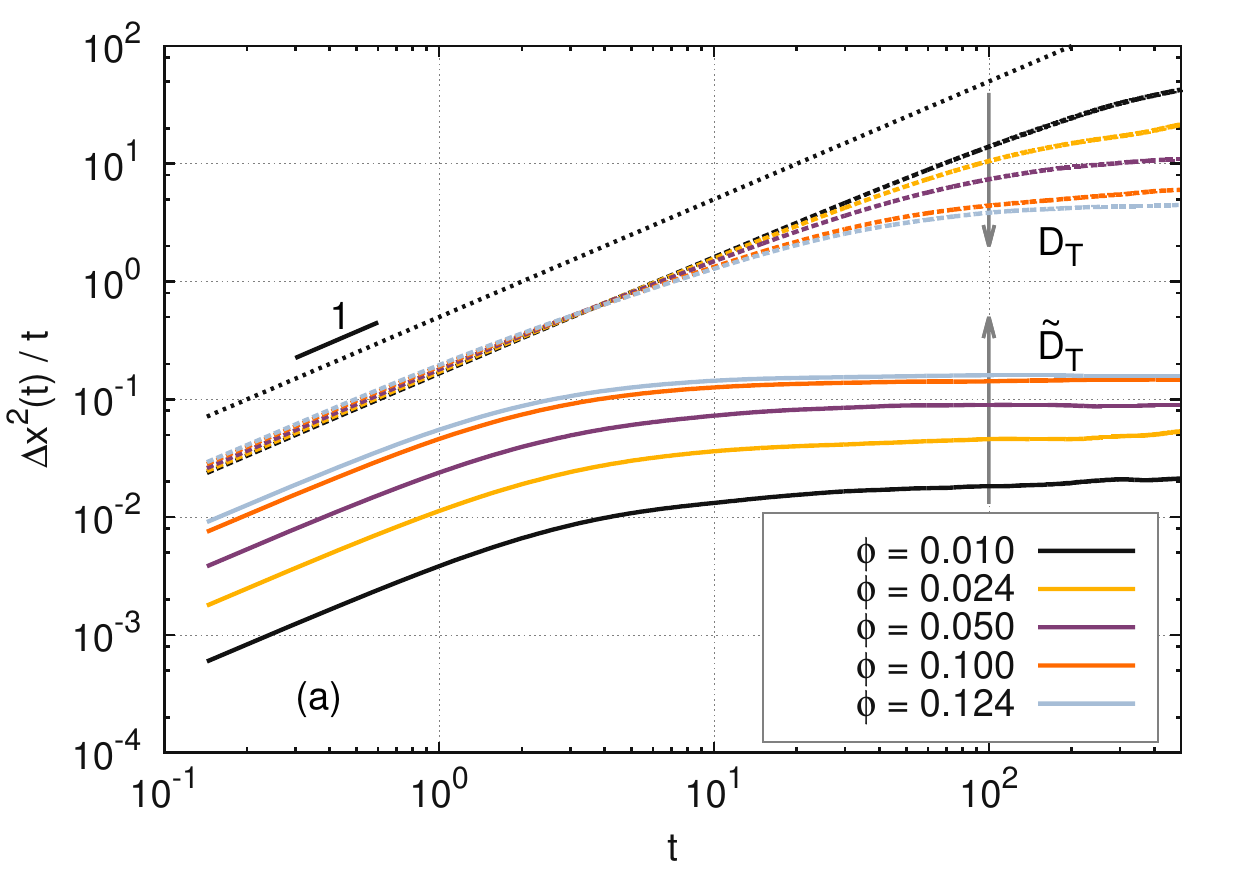}\\
  \includegraphics[width=0.45\textwidth]{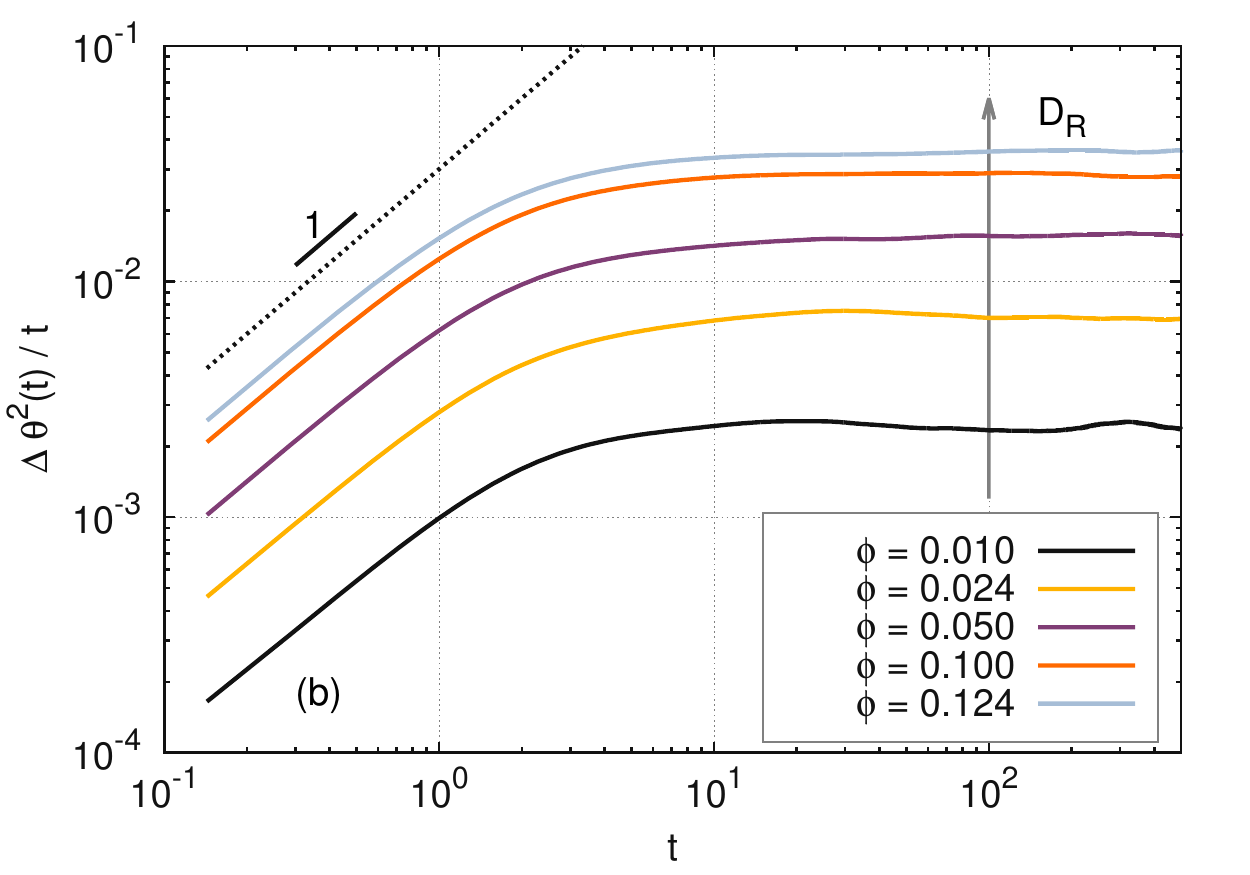}
  \caption{\label{f:diff_single}(color online) Translational and
    rotational diffusivities for a system of pullers ($\alpha=+2$) at
    various concentrations. (a) Translational diffusivities obtained
    from the mean-squared displacements in the fixed-lab reference
    frame $D_T(t)$, and the body frame $\body{D}_T(t)$ (with the
    displacement due to the intrinsic swimming motion suitably
    removed), dashed and solid lines, respectively; and (b) the
    rotational diffusivities $D_R(t)$. Lines of slope one,
    corresponding to purely ballistic motion, have been drawn for
    comparison. The arrows show the increase/decrease of the
    various diffusivities as a function of concentration $\phi$.}
\end{figure}

The diffusivities $D_T(t)$ and $\body{D}_T(t)$, defined in
Eqs.~\eqref{e:diff}~and~\eqref{e:diff_body}, for a system of pullers
($\alpha=+2$), at various concentrations, are shown in
Figure~\ref{f:diff_single}. The standard diffusivities $D_T(t)$ show a
ballistic regime which extends to very long times $t\simeq 100$, after
which the slope starts to decrease, indicating a transition towards a
diffusive regime. However, purely diffusive motion (slope of zero) is
only obtained for the highest concentrations and alpha values. This
behavior has been analyzed in detail
in~ref.~\citenum{Ishikawa:2007cf}. In contrast, $\body{D}_T(t)$
reaches the diffusive regime at much shorter times, for all
concentrations and all non-zero values of alpha. We note that both
quantities are measuring related phenomena, but the latter
does so directly, since the motion due to the inherent particle
swimming has been removed from the analysis. As a function of
concentration, $\body{D}_T$ shows a clear increase, which is
consistent with the interpretation of the diffusive motion being
caused by interactions with neighboring particles: as the
concentration increases, so does the number of neighbors, and thus the
strength of the interactions. In contrast, we see that $D_T$ decreases
with concentration. The reason for this is simple, since $\text{D}_T$
mainly measures the effect of the particles' own swimming, an increase
in the diffusive behavior can only hinder this motion. The rotational
diffusivities $D_R$, also shown in Figure~\ref{f:diff_single}, present
the same basic features as $\body{D}_T$, the onset of the diffusive
regime is obtained within the same time interval and they exhibit a
similar concentration dependence. Apart from a difference in scale,
there is no clear differentiating factor between these two quantities.
Similar results are obtained for all other non-zero values of $\alpha$
considered; where, as was pointed out by Ishikawa and
Pedley\cite{Ishikawa:2007cf}, the effect of increasing (decreasing)
$\alpha$ is the same as that of increasing (decreasing) the
concentration, at least with respect to the diffusive motion of the
squirmers, since they both lead to an increase (decrease) in the
strength of the hydrodynamic interactions among particles.
\subsection{Velocity fluctuations and correlations}
\begin{figure}[ht!]
  \centering
  \includegraphics[width=0.5\textwidth]{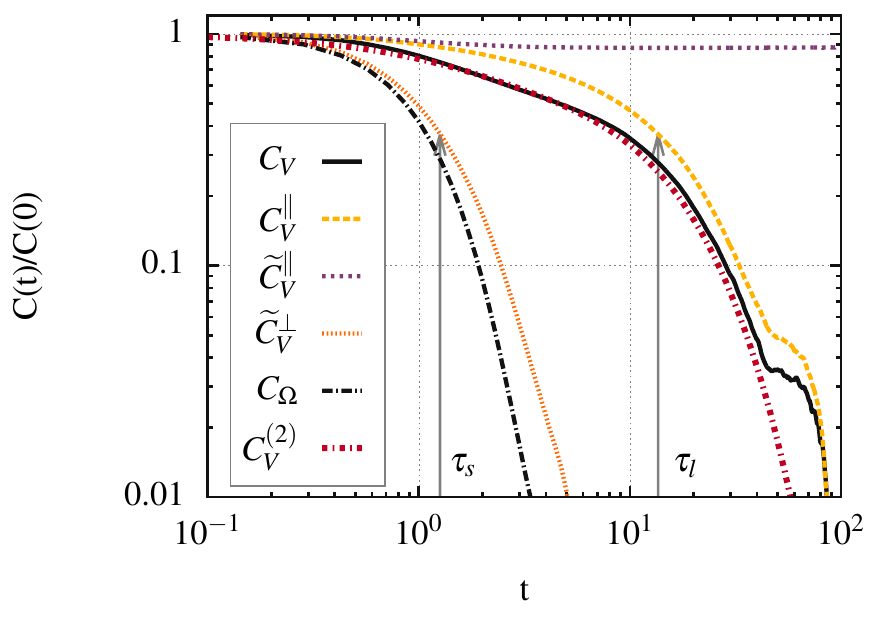}
  \caption{\label{f:vcf_a2}(color online) Single-particle velocity
    time-correlation functions for a system of pullers ($\alpha=+2$)
    at $\phi=0.124$. The two-time relaxation approximation to the
    velocity correlation function $C_V^{(2)}(t)/C_V(0)$
    (Eq.~\ref{e:cv2}), with the amplitudes and relaxation times
    obtained from the simulation, is also shown (dash-dot red line).}
\end{figure}
To have a better understanding of the how the diffusive motion arises,
we now consider the single particle velocity correlation functions
defined in Eqs.\eqref{e:cvt}~-~\eqref{e:cvt_XY}. We present in detail
the results obtained for a system of pullers ($\alpha = +2$) at $\phi
= 0.124$, shown in Figure~\ref{f:vcf_a2}, but the same analysis
applies for the other values of $\phi$ ($\alpha$) we have studied. The
standard definition of the velocity auto-correlation $C_V(t)$ results
in a function that shows a two-time decay process, as evidenced by the
shoulder that appears around $1 < t < 10$. The initial decay ($t\simeq
1$) is due to the changing configuration of neighboring particles (and
their self-generated flow field), while the second, long time decay
($t\simeq 10$) arises due to the reorientation of the particles'
swimming direction, which is a consequence of the particle-particle
collisions. Confirmation for this interpretation is provided by the
three other correlation functions we have defined. Consider first the
decay in correlations of the parallel velocity components,
$C_V^\parallel(t)$ (lab frame) and $\body{C}_V^\parallel$ (particle
frame). Both functions show the same initial decay, caused by local
fluctuations in the velocity field, as the particle has not
experienced any significant change in its swimming direction. However,
at longer times, these two functions show a completely different
behavior. The parallel correlations measured within the fixed lab
system $C_V^\parallel(t)$ exhibit a slow (long-time) decay before
eventually collapsing onto the full correlation function $C_V(t)$;
while the correlations measured within the particle's frame of
reference reach a plateau after the initial decay. This means that
while the particle is constantly reacting to the flow field generated
by its neighbors, by changing it's swimming speed, these fluctuations
decay very fast, typically within the time it takes for the particle
to travel its diameter (the characteristic length over which the flow
field varies significantly). As such, the long-time decay in both
$C_V(t)$ and $\body{C}_V^\parallel(t)$ is due primarily to changes in
the orientation of the swimming direction, and not to changes in the
magnitude of the swimming speed. As expected, the correlations in the
perpendicular velocities $\body{C}_V^\perp(t)$ show a very fast decay,
over a characteristic time equal to that of the initial short-time
decay of the parallel velocity components. We also see a clear
correspondence between the decay in correlations of the angular
velocity $C_{\Omega}(t)$ and the perpendicular velocity
$\body{C}_V^\perp(t)$. 

Finally, the characteristic times $\tau_s$ and $\tau_l$ for these two
processes, governed by the hydrodynamic interactions and the particle
collisions, can be obtained by assuming an exponential decay
$\exp{\left(-t/\tau\right)}$ for the appropriate correlation function:
$\body{C}_V^\perp(t)$ and $C_V^\parallel(t)$ for $\tau_s$ and
$\tau_l$, respectively (see Figure~\ref{f:vcf_a2}). The two-time scale
relaxation of the velocity fluctuations can then be approximated as
\begin{align}
  C_V^{(2)}(t;\phi,\alpha) &= \chi_s(\phi,\alpha) e^{-t/\tau_s(\phi,\alpha)}
  + \chi_l(\phi,\alpha)e^{-t/\tau_l(\phi,\alpha)} \label{e:cv2}\\
  \chi_s &= \avg{\delta\vec{V}\cdot\delta\vec{V}} \notag\\
  \chi_l &= U_\parallel^2 \notag
\end{align}
with $U_\parallel=\avg{\vec{V}\cdot\uvec{e}_3}$ and where we have
explicitly shown the dependence of the amplitudes $\chi$ and decay
times $\tau$ on the concentration and swimming mode of the particles.
Figure~\ref{f:vcf_a2} shows excellent agreement between this two-time
relaxation process (with all four parameters obtained from the
simulation data) and the full correlation function $C_V(t)$ defined in
Eq~\eqref{e:cvt}. Furthermore, in what follows we show how the long-time decay can be
simplified and interpreted in terms of a binary collision process
between the swimmers. 
\subsection{Scaling}
To confirm the interpretation we have given for the diffusive motion
of the squirmers, and the emergence of the two time-scales, we perform
a scaling analysis for the relevant parameters (diffusion coefficients
and collision time scales), as a function of concentration $\phi$.
Although a similar analysis has been presented
in~ref.~\citenum{Ishikawa:2007cf} by Ishikawa and Pedley, they have
not considered the \textit{effective} diffusion $\body{D}_T$, which
means that only the long-time behavior of the system, given by the
particle collisions, was analyzed (the hydrodynamic interactions which
give rise to the rapidly decaying velocity fluctuations are completely
masked by the swimming motion). In a subsequent
study\cite{Ishikawa:2010ho}, the authors considered the diffusion of
fluid particles, as well as inert colloidal particles, in a suspension
of swimmers, and in this case they were able to observe the emergence
of a second (shorter) time-scale. Our results, which are in agreement
with their scaling arguments, provide a complementary view of the
diffusive motion of these squirmer systems. The benefit of the
analysis we propose lies in the fact that all relevant time and length
scales can be obtained just from the motion of the squirmers, i.e.,
there is no need to consider the motion of the fluid or to introduce
inert (non-swimming) particles. This results in simpler simulations
and will also be relevant when trying to compare with experimental
data.
\begin{figure}[ht!]
  \centering
  \includegraphics[width=0.45\textwidth]{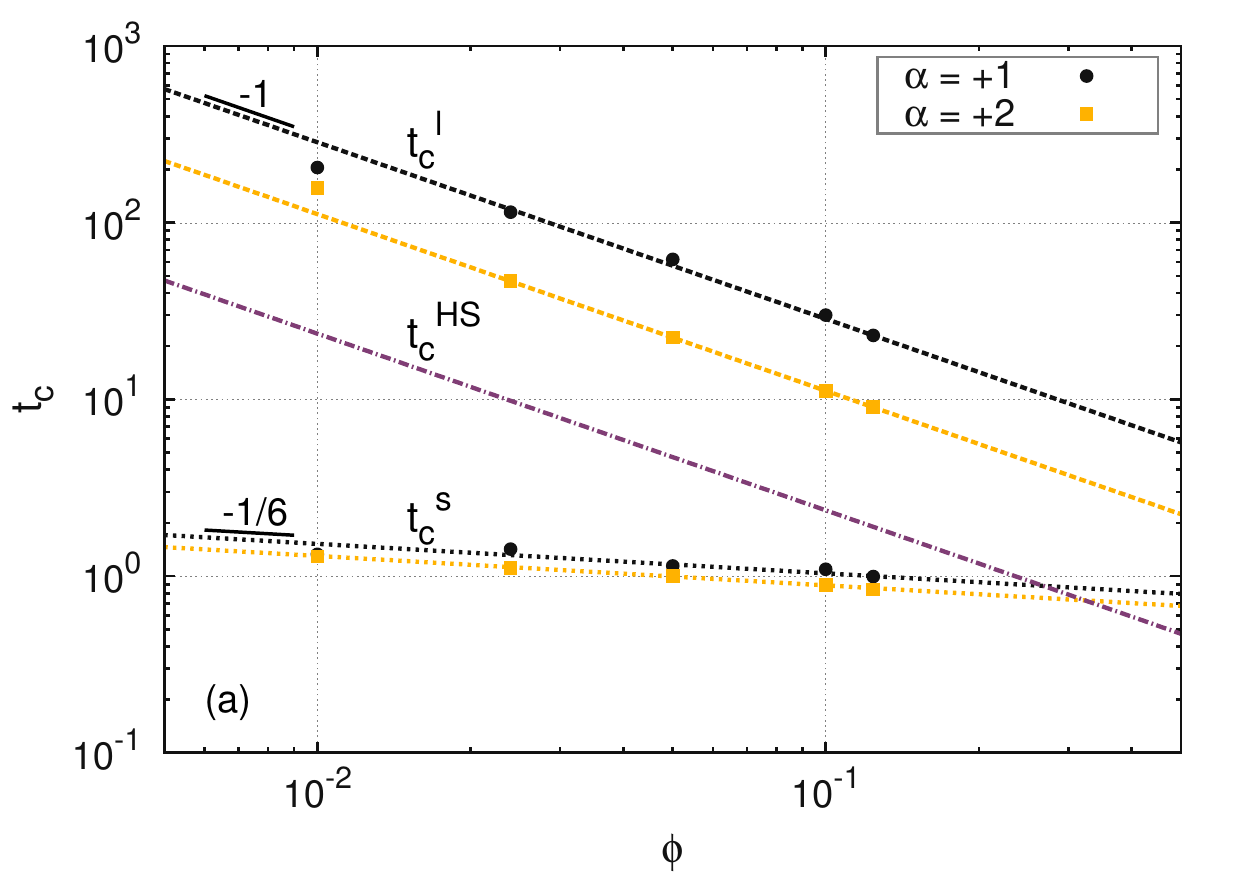}\\
  \includegraphics[width=0.45\textwidth]{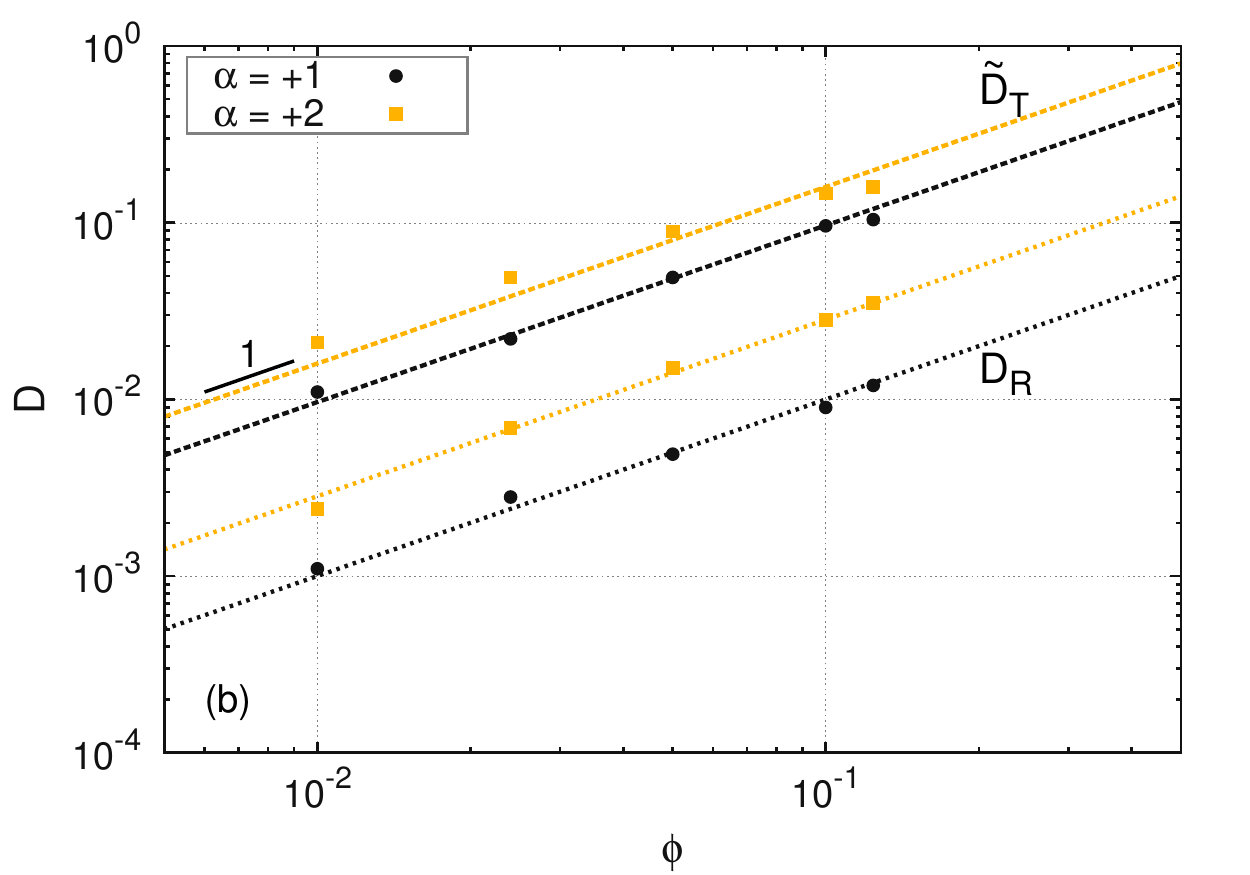}\\
  \includegraphics[width=0.45\textwidth]{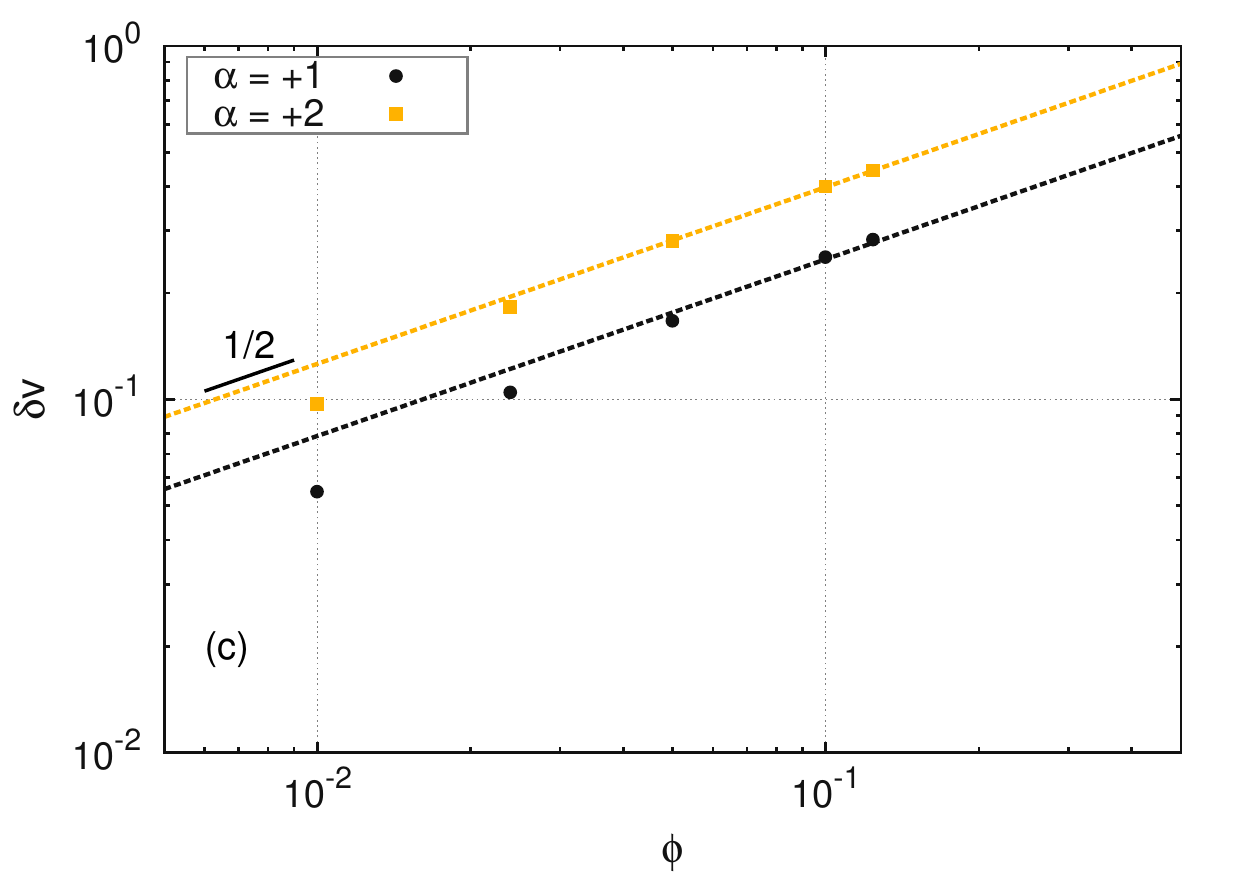}
  \caption{\label{f:tau_scale}(color online) (a) Scaling of the
    ``collision'' times, $t_c^l$ and $t_c^s$, with concentration
    $\phi$. Also shown is the collision time $t_{c}^{\text{HS}}$ for
    an equivalent system of hard spheres, with a (scaled) velocity of
    $U_c=1$, given by the kinetic theory of gases. (b) Scaling of
    the translational $\body{D}_T$ and rotational $D_R$ diffusion
    coefficients with concentration. (c) Scaling of the velocity
    fluctuations with concentration.}
\end{figure}

\subsubsection{Correlation times~~}First, lest us consider the
long-time dynamics of the system, which is determined by the
particle-particle collisions of the swimmers. From the kinetic theory
of gases\cite{McQuarrie}, we know that the mean free path $\lambda$
(or average distance between collisions) should be proportional to the
inverse of the concentration of particles. In reduced units, we have
\begin{equation}\label{e:lambda_mfp}
  \lambda = \frac{2\sqrt{2}}{3} \left(\sigma_c^2\,\phi\right)^{-1}
\end{equation}
with $\phi$ the packing fraction, and $\sigma_c$ the ``collision'' or
cross-section diameter, which is not necessarily equal to the actual
physical diameter of the particle. This difference is due to the
self-generated flow profile around each particle, and as such should
only depend on the swimming mode $\alpha$, not on the concentration of
particles. For the moment though, let us assume that we are dealing with
hard-spheres $\sigma_c = 2$, the average time between collisions is
simply $t_c^{\text{HS}} = \lambda^{\text{HS}} / U_c$, which gives
\begin{equation}\label{e:tau_mfp}
  t_c^{\text{HS}} = \frac{\sqrt{2}}{6}\left(U_c\,\phi\right)^{-1}
\end{equation}
where $U_c$ is the average velocity of the particles. In thermal
systems, this velocity would be determined by the reservoir and be
independent of concentration, so that one obtains the well known
$\phi^{-1}$ dependence for the collision time $t_c^{\text{HS}}\propto
\phi^{-1}$. Although the velocity of our squirmers is not
concentration independent, to first order, we can safely assume that
the particles are all swimming at an average velocity which is near
the swimming velocity of an isolated squirmer, $U_c=1$. Therefore, the
collision time for our systems should show the same concentration
dependence predicted by kinetic theory, $t_c\propto \phi^{-1}$. To
obtain the collision times $t_c$ from the decay times $\tau$ computed
from the simulations, we use the Enskog
approximation\cite{McQuarrie,Hansen}
\begin{equation}
  t_c = 2\tau/3
\end{equation}
Our results, shown in Figure~\ref{f:tau_scale}, confirm the scaling
predictions for the long-time decay $t_c^{l}=2\tau_l/3$. Of special
interest is the difference between $t_c^{l}$ for the different
squirming modes, with the collision time decreasing with increasing
$\alpha$. This can be explained by an increase in the effective size
of the particle, as the mean free path $\lambda$ is inversely
proportional to the collision diameter $\sigma_c$. We also show the
mean collision time $t_c^{\text{HS}}$ for an equivalent system of
hard-spheres. The fact that the collision times for the squirmers are
significantly larger than the corresponding hard-spheres values is
surprising, since it implies that the squirmers have a cross-section
diameter which is smaller than the hard-sphere diameter of the particle
($t_c\propto\sigma_c^{-2}$). The reason for this lies in the collision
\textit{dynamics} of the squirmers.

The scaling of the fast time-scale $\tau_s$ (or $t_c^{s}$) is harder
to elucidate, since it shows only a very weak concentration
dependence. Given that $\tau_s$ is related to the velocity
fluctuations caused by the short-range hydrodynamic interactions
between particles, it should be related to the time it takes for the
flow region around a particle to change: this is essentially the time
necessary for a particle to swim its diameter $t\simeq 2$. Our
results, also shown in Figure~\ref{f:tau_scale}, agree with this rough
estimate and also indicate a power-law behavior with an exponent of
$\simeq -1/6$. Although the decrease with concentration seems clear,
it is very difficult to accurately measure such small variations and
we have no suitable explanation for the value of this exponent (given
by a fit to the data). These results are analogous to those obtained
by Ishikawa et al.\cite{Ishikawa:2010ho} for the time-scale of
diffusing \textit{fluid} particles in a suspension of squirmers.

\subsubsection{Diffusion coefficients~~}For the scaling of the
translational diffusion coefficients, a simple dimensional analysis
yields
\begin{equation}\label{e:d_scale}
  D_T \propto U_c^2\, t_c
\end{equation}
where $U_c$ and $t_c$ are the characteristic velocity and time scales
of the ``collision'' process which gives rise to the diffusive
motion\cite{Ishikawa:2010ho}. For the standard diffusion coefficient
$D_T$ (Eq.~\eqref{e:DDT}), the reorientation of the particles is
caused by the particle-particle collisions, thus $t_c = t_c^l$ and
$D_T \propto \phi^{-1}$ (assuming a constant swimming velocity). This
behavior has been analyzed in detail
in~ref.~\citenum{Ishikawa:2007cf}. More relevant to our study is the
scaling of the \textit{effective} diffusion coefficient $\body{D}_T$
(Eq.~\eqref{e:EDT}), in which the motion due to the particle swimming
has been removed. Our results, shown in Figure~\ref{f:tau_scale},
indicate a linear dependence with concentration $\body{D}_T\propto
\phi$. This is precisely the scaling behavior reported by Ishikawa et
al.\cite{Ishikawa:2010ho} for the diffusion of inert/fluid particles,
and the same concentration dependence that has been observed
experimentally for the diffusion of tracer particles in swimming
suspensions (for both pushers and
pullers)\cite{Kim:2004bq,Leptos:2009kd}. This is not surprising, since
we can consider that this diffusive motion arises due to collisions
between the particle (swimmer or not) and the fluid. By definition, it
is clear that the velocity and time scales cannot be the same swimming
speed or collision time used to define $D_T$. In this case, the
characteristic velocity scale will be set by the velocity fluctuations
$\delta v$ around the average swimming velocity, which can be directly
measured by computing the velocity components perpendicular to the
swimming axis of the particles $V^\perp$. For these fluctuations, our
simulations indicate a square-root dependence with concentration
$\delta v\propto \sqrt{\phi}$ (see Figure~\ref{f:tau_scale}). The
characteristic time-scale for this process will be the time necessary
for the fluid flow surrounding a given particle to show considerable
variations, and this can be estimated by the time it takes the
particle to travel its own diameter $t_c\propto 1$. We thus recover
the linear dependence given by the simulations. The same scaling
behavior is obtained for the rotational diffusion coefficient
$D_R\propto \phi$.
\subsubsection{Collision diameters~~}
\begin{figure}[ht!]
  \includegraphics[width=0.45\textwidth]{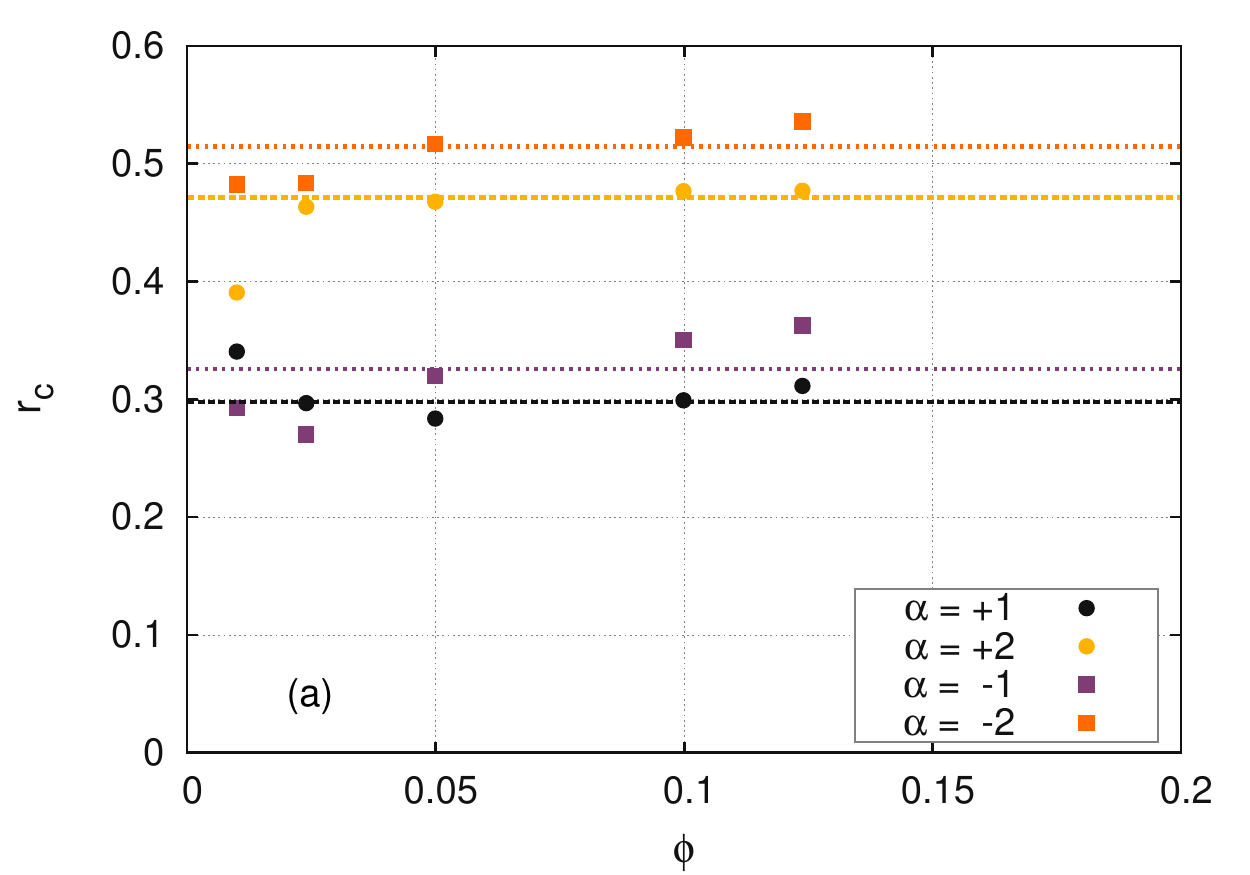}\\
  \includegraphics[width=0.45\textwidth]{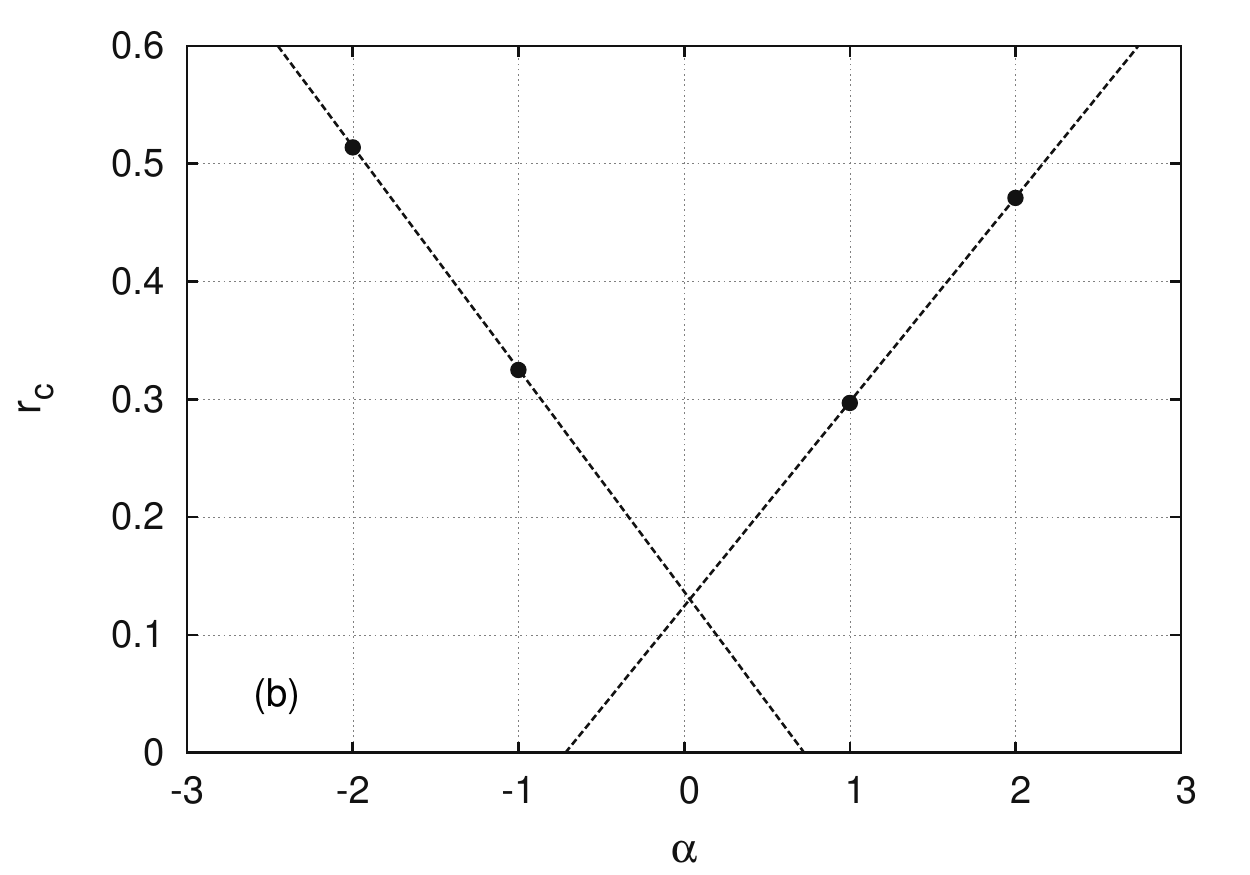}
  \caption{\label{f:sigma_scale}(color online) Effective collision
    radius $r_c$ of the squirmers as a function of concentration
    $\phi$. Results were obtained by using the mean free path
    expression for a system of hard-spheres given by the kinetic
    theory of gases, with the collision time given by $\tau_\lambda =
    \tau_l/16$ ($\tau_l$ is the decay time in the velocity
    correlations given by particle-particle collisions).}
\end{figure}

So far, we have not considered
the effect of the squirming mode $\alpha$ on the diffusive properties
of the swimmers. While it is clear that increasing the magnitude of
$\alpha$ will give rise to stronger hydrodynamic interactions, and
thus increase the diffusion of the particles (as can be seen in
Figure~\ref{f:tau_scale}), it is not clear how this dependence can be
quantified. We propose that $\alpha$, which defines the strength/range
of the self-generated fluid-flow around a squirmer
(Eq.~\eqref{e:squirm_field}), can be directly related to the effective
``collision'' diameter of the swimmers. From the expression of the
mean free path, Eq.~\eqref{e:lambda_mfp}, we obtain the following for
the collision radius $r_c = \sigma_c / 2$
\begin{equation}\label{e:a_mfp}
  r_c = \left(3\sqrt{2}\,U_c\,t_c\,\phi\right)^{-1/2}
\end{equation}
Assuming $t_c= f(\alpha)\,\phi^{-1}$, with $f(\alpha)$ a function
only of the squirming mode $\alpha$, and considering the swimming
velocity to be constant ($U_c = 1$), we arrive at a concentration
independent collision radius
\begin{equation}
  r_c \propto  \left[f(\alpha)\right]^{-1/2}
\end{equation}
As $f(\alpha)$ is a decreasing function of $\alpha$, the effective
collision radius of the particle should increase with the magnitude of
the squirming mode. Our results for the collision radius, given by
Eq.~\eqref{e:a_mfp}, with $U_c=U_\parallel=\avg{\vec{V}\cdot\uvec{e}_3}$ and $t_c = 2/3
\tau^l$ obtained directly from the simulations, are shown in
Figure~\ref{f:sigma_scale}. Allowing for the large uncertainties in
measuring the decay times at low concentrations, we obtain very good
agreement with the previous scaling analysis: the radii show only a
small variation with $\phi$, and a clear distinction is observed as
$\lvert\alpha\lvert$ is increased. We note however that the pushers
seem to present a larger collision radius than the pullers, something
which is difficult to explain with the simple collision model we have
presented. 

Also shown in Figure~\ref{f:sigma_scale} is the mean
collision radius $r_c$ as a function of $\alpha$. Although more data
is needed to accurately specify the functional form of $f(\alpha)$, a
simple linear dependence predicts a value of $r_c(\alpha=0)$ which is
an order of magnitude smaller than the hard-sphere radius of the
particles $r_c\simeq 0.1$. This is consistent with our simulations,
from which we were unable to obtain reliable estimates for $\tau_l$,
because the velocity correlation functions exhibited no substantial
decay over the time scales we studied $t\simeq 10^2\sim10^3$. However,
we were able to obtain estimates for the effective diffusion
coefficients $\body{D}_T$ of these neutral squirmers, which are an order
of magnitude smaller than the corresponding values for $\alpha=\pm 1$.
This means that the difference in collision times, with respect to a
corresponding system of hard-spheres (see
Figure~\ref{f:tau_scale}), will be even larger for the neutral squirmers
than it is for the swimmers with non-zero $\alpha$ values. This is
consistent with an interpretation in terms of hard-spheres when
analyzing the differences among squirmers (larger $\alpha$ equivalent
to larger collision radius $r_c$), but it can appear contradictory
when comparing to actual hard-sphere systems. After all, we could
expect the systems for $\alpha=0$, in which the flow profile due to
the squirming motion is most localized ($U(r)\propto r^{-3}$), to be
closest to a system of hard-spheres. Yet our results indicate that the
collision radius of the neutral squirmers is an order of magnitude
smaller than the actual radius of the particles. As we have already
mentioned, this is due to the difference in the collision dynamics of
the particles, which differ dramatically from that of
hard-spheres\cite{Ishikawa:2006hf,Llopis:2010in}. Although the exact
collision process will depend not only on the relative velocity of the
particles, but also on their relative orientations, the squirming
motion results in deflection angles which are (on average) smaller
than the corresponding hard-sphere values. In essence, this means that
the flow fields generated by the particles allows them to swim past
each other with a relatively small change in their swimming direction
(again, as compared to with an actual hard-sphere collision).

Finally, using Eq.~\eqref{e:a_mfp}, the velocity correlation function
for a suspension of squirmers (Eq.~\eqref{e:cv2}) can be reduced to
the following functional form
\begin{equation}
  C_V(t;\phi,\alpha) = \chi_s\, e^{-t/\tau_s}
  + U_\parallel^2 e^{-\sqrt{2} \,\sigma_c\, U_\parallel \phi\,t} 
\end{equation}
where the long-time decay process is expressed only in terms of the
average swimming speed $U_\parallel(\phi,\alpha)$ and the
\textit{hard-sphere} collision diameter $\sigma_c(\alpha)$ of the
squirmers. The former has a weak linear dependence on concentration
$U_\parallel = 1 - A(\alpha)\phi$, while the latter is concentration
independent, as was shown in Fig.~\ref{f:sigma_scale}. The scaling
behavior for the short time process is still not understood, but our
simulation results suggest a power law behavior for the concentration
dependence of the amplitude $\chi_s$ and decay time $\tau_s$ of the
velocity fluctuations, with an exponent of $1/2$ and $-1/6$,
respectively; i.e., $\chi_s\propto \phi$ and $\tau_s\propto
\phi^{-1/6}$. A more detailed analysis is required to firmly establish
these scaling relationships, as well as to determine the dependence on
the swimming parameter $\alpha$. Work along these lines is in
progress.

\section{Conclusions}
We have investigated the hydrodynamic interactions of suspensions of
squirmers using a modified version of the smoothed profile method (SP)
for particle dispersions. The SP method allows one to fully resolve the
hydrodynamic interactions in many particle dispersions in an accurate
and efficient manner, and we have shown how it can be extended to
systems with slip boundary conditions, such that it is possible to
describe squirmers (active swimmers which move due to self-generated
surface tangential velocities). The validity of the method was
confirmed by comparing the simulation data with exact results for
the case of a single swimmer, for which we recover the correct
swimming speed and are able to accurately reproduce the fluid flow
generated by the squirming motion, and for two aligned swimmers at a
fixed distance, for which we recover the expected hydrodynamic force.
The advantage of the SP method for swimming particles, as opposed to
Stokesian Dynamics (which has been successfully, and extensively, used
to study these systems) is its applicability to particle dispersions in
complex fluids. This is relevant in the case of swimming
micro-organisms, since the role of the nutrient and the possibility of
having a non-Newtonian host fluid must be considered when comparing
with experiments.

In this paper we have analyzed the effect of the hydrodynamic
interactions on the motion of semi-dilute squirmer suspensions, up to
volume fractions of $\phi=0.124$, for various swimming modes $|\alpha|
\le 2$. Although we have no yet included thermal fluctuations in our
description, the swimming motion of the particles gives rise, over
sufficiently long time scales, to a diffusive regime. In order to
distinguish between the contributions due to the hydrodynamic
interactions, caused by the squirming motion, and those due to the
particle-particle collisions, which are the two basic mechanisms
responsible for the diffusive motion, we have analyzed the particle
dynamics in terms of movement due to the inherent swimming of the
particles, and that due to the (hydrodynamic) interactions among them.
This is easily done by looking at the motion from the particle's own
frame of reference, i.e., decomposing the motion parallel and
perpendicular to the swimming axis. This analysis has allowed us to
demonstrate the appearance of two distinct time scales within our
system, one related to the time between particle-particle collisions,
the other to the fluid-particle interactions. This two-time scale
nature of the particle interactions in swimming suspensions can be
clearly seen in the two-time relaxation of the velocity correlation
function. We are thus able to define an effective hydrodynamic
diffusion coefficient (corresponding to the short-time fluid/particle
interactions), in which the self-motion of the particle has been
removed, which shows a linear scaling with concentration. In contrast,
the standard diffusion coefficient is inversely proportional to the
concentration of swimmers. This is in agreement with simulation and
experimental results on \textit{tracer} diffusion in swimming
suspensions. Additionally, since the long-time dynamics of the system
is related to the particle-particle collisions, we have used the
well-known results from kinetic theory to deduce an effective,
concentration independent, collision radius for our swimmers. Due to
the complex collision dynamics of these particles, this collision
radius is actually smaller than the hard-sphere radius of the
particle, and increases with increasing $\alpha$.
\section*{Acknowledgements}
The authors would like to express their gratitude to the Japan Society
for the Promotion of Science for financial support (Grants-in-Aid for
Scientific Research KAKENHI No. 23244087).
\bibliography{active,kapsel,books}
\end{document}